\PassOptionsToPackage{table,dvipsnames}{xcolor}
\documentclass[sigconf, nonacm, screen]{acmart}
\AtBeginDocument{%
  }

\setcopyright{none}

\usepackage{amsmath,amsfonts}
\usepackage{graphicx}
\usepackage{textcomp}
\definecolor{salmon}{rgb}{1.0, 0.55, 0.41}
\usepackage{multirow}
\usepackage{threeparttable}
\usepackage{cancel}

\definecolor{salmon}{rgb}{1.0, 0.55, 0.41}

\usepackage{paralist}
\newcommand{\compactpara}[1]{\smallskip\noindent\textbf{#1:}}

\usepackage{tikz}
\usepackage{amsmath}
\usepackage{todonotes}
\usepackage{amsfonts}
\usepackage{listings}
\usepackage{xcolor}
\usepackage{soul} \setul{}{1pt}
\usepackage{subscript}
\usepackage{mathtools}
\usepackage[noend,ruled,linesnumbered]{algorithm2e}
\usepackage{bussproofs}
\usepackage{multicol}

\usepackage{cleveref}
\usepackage{tcolorbox}
\usepackage[inline]{enumitem}
\usepackage{adjustbox}
\usepackage{makecell}
\usepackage[T1]{fontenc}
\usepackage{stfloats}
\usepackage{pbalance}

\newcommand{\redul}[1]{\setulcolor{OrangeRed}\ul{#1}}
\newcommand{\blueul}[1]{\setulcolor{NavyBlue}\ul{#1}}
\newcommand{\greenul}[1]{\setulcolor{ForestGreen}\ul{#1}}
\newcommand{\orangeul}[1]{\setulcolor{Dandelion}\ul{#1}}

\newcommand{\toolx}{\tool{}\textsc{-x86}\xspace}
\newcommand{\toolarm}{\tool{}\textsc{-arm}\xspace}

\newtheorem{step}{Step}

\begin{document}

\title{Automated Template-free Synthesis of Instruction-Centric Leakage Contracts for Black-Box CPUs
}

\newcommand{\boldparagraph}[1]{\medskip\noindent\textbf{#1:}}

\newcommand{\tool}{\textsc{malcos}\xspace}
\newcommand{\checker}{{\texttt{Checker}}\xspace}
\newcommand{\refiner}{{\texttt{Refiner}}\xspace}
\newcommand{\postprocessor}{\texttt{Post\-pro\-ces\-sor}\xspace}
\newcommand{\malcos}{\tool} %
\newcommand{\revizor}{\textsc{Revizor}\xspace}
\newcommand{\scamv}{\textsc{Scam-V}\xspace}
\newcommand{\muasm}{\ensuremath{\mu}ASM\xspace}
\newcommand{\spectre}{\textsc{Spectre v1}\xspace}
\newcommand{\langName}{\textsc{icl}\xspace}
\newcommand{\icl}{\langName}
\newcommand{\icentric}{instruction-centric}

\newcommand{\isdef}[0]{\ensuremath{\mathrel{\overset{\makebox[0pt]{\mbox{\normalfont\tiny\sffamily def}}}{=}}}}

\newcommand{\mtt}[1]{\ensuremath{\mathtt{#1}}}
\newcommand{\funname}[1]{\mtt{#1}}
\newcommand{\fun}[2]{\ensuremath{{{\funname{#1}\left(#2\right)}}}\xspace}
\newcommand{\ctrace}[1]{\fun{CTR}{#1}}
\newcommand{\htrace}[1]{\fun{HTR}{#1}}
\newcommand{\csat}[2]{\ensuremath{#1 \vdash #2}}
\newcommand{\ceq}{\ensuremath{=_C}}
\newcommand{\ceqgen}[1]{\ensuremath{=_{#1}}}
\newcommand{\pceq}[3]{\ensuremath{#1 \vdash #2 \ceq #3}}
\newcommand{\pceqgen}[4]{\ensuremath{#1 \vdash #2 \ceqgen{#4} #3}}

\date{}

\author{Elvira Moreno Sánchez}
\authornote{Both authors contributed equally to this research.}
\affiliation{%
  \institution{IMDEA Software Institute}
  \institution{Universidad Politécnica de Madrid}
  \country{}
}
\email{elvira.moreno@imdea.org}

\author{Tiziano Marinaro}
\authornotemark[1]
\affiliation{%
  \institution{CISPA Helmholtz Center for Information Security}
  \institution{Saarland University}
  \country{}
}
\email{tiziano.marinaro@cispa.de}
\author{Ryan Williams}
\affiliation{%
  \institution{Northeastern University}
  \country{}
}
\email{williams.ry@northeastern.edu}

\author{Marco Patrignani}
\affiliation{%
  \institution{University of Trento}
  \country{}
}
\email{marco.patrignani@unitn.it}

\author{Roberto Guanciale}
\affiliation{%
  \institution{KTH Royal Institute of Technology}
  \country{}
}
\email{roberto.guanciale@kth.se}

\author{Hamed Nemati}
\affiliation{%
  \institution{KTH Royal Institute of Technology}
  \country{}
}
\email{hamed.nemati@kth.se}

\author{Marco Guarnieri}
\affiliation{%
  \institution{IMDEA Software Institute}
  \country{}
}
\email{marco.guarnieri@imdea.org}

\renewcommand{\shortauthors}{E. Moreno Sánchez and T. Marinaro, et al.}

\Crefformat{section}{\S#2#1#3}
\crefformat{section}{\S#2#1#3}
\Crefformat{subsection}{\S#2#1#3}
\crefformat{subsection}{\S#2#1#3}
\Crefformat{subsubsection}{\S#2#1#3}
\crefformat{subsubsection}{\S#2#1#3}

\crefrangeformat{section}{\S#3#1#4--#5#2#6}
\Crefrangeformat{section}{\S#3#1#4--#5#2#6}
\crefrangeformat{subsection}{\S#3#1#4--#5#2#6}
\Crefrangeformat{subsection}{\S#3#1#4--#5#2#6}
\crefrangeformat{subsubsection}{\S#3#1#4--#5#2#6}
\Crefrangeformat{subsubsection}{\S#3#1#4--#5#2#6}

\begin{abstract}

Side-channel attacks pose a significant security threat for modern computing platforms, because they exploit subtle discrepancies in CPU behaviors to leak sensitive information.
To model the information leaked by a CPU via microarchitectural side-channels, recent work proposed \textit{leakage contracts}: an ISA-level security abstraction that provides the foundations for secure CPU programming.
Unfortunately, due to the complexity of current microarchitectures, 
devising a leakage contract for a  CPU requires extensive manual effort and thus modern CPUs lack dedicated leakage contracts.

We present a methodology to extract instruction-centric leakage contracts for major CPU architectures with minimal manual intervention.
We implemented this technique in \tool{}, the first template-free tool that automates the  synthesis of leakage contracts for black-box CPUs.
We evaluate \tool on  {\tt x86} and {\tt ARM}  CPUs, and show that the contracts it synthesizes are precise and sound with respect to all leaks observed during synthesis.
Our results demonstrate that learning leakage contracts from black-box CPUs is feasible. %

\end{abstract}


\maketitle

\section{Introduction}\label{sec:intro}

Side-channel attacks exploit variations in processor behavior---such as execution time~\cite{evicttime,reloadrefresh,flushreload,primeprobe,flushflush} or cache access patterns~\cite{Aciicmez:2006:TCA:2092880.2092891,Neve:2006:AAC:1756516.1756531,Tsunoo03cryptanalysisof,gullasch2011cache}---to infer sensitive information from otherwise secure software.
To safeguard security-critical software, 
developers often adopt the Constant Time 
model~\cite{DBLP:conf/pkc/Bernstein06}.
This model 
assumes that the only sources of side-channel leaks are control-flow instructions and memory accesses, and secure programs need to make both  secret-independent to prevent leaks. %
This assumption, however, is violated in modern CPUs. 
Processor optimizations---ranging from arithmetic optimizations for values like 0 and 1~\cite{atoofian2005improving} to advanced speculative execution techniques~\cite{lipp2018meltdown,kocher2020spectre}---can introduce new microarchitectural side-channel vulnerabilities.
Hence, there is a growing need for techniques that enable developers and system architects to precisely characterize and mitigate such leaks.\looseness=-1

\emph{Leakage contracts}~\cite{guarnieri2021hardware,scamv} augment the Instruction Set Architecture (ISA) with a specification of all observable side-channel leaks  within a  CPU.
This enables secure system development since programmers are made aware of the exploitable side-channels that are traditionally obscured at the ISA level.
Unfortunately, constructing leakage contracts for modern CPUs is a complex task: it requires extensive reverse engineering, expert knowledge, and significant time investment~\cite{hofmann2023speculation}, making it impractical to apply across the diverse landscape of commercially available CPUs.

Recently, several approaches have been proposed to synthesize \emph{instruction-centric} contracts~\cite{wang2025synthesis, hsiao2024rtl2m, dinesh2025h, DBLP:conf/date/MohrG024, DBLP:conf/sp/DineshPF24}, i.e., a specific class of contracts where leaks are characterized as a function of instruction operands, which can capture subtle data-dependent instruction-level optimizations~\cite{vicarte2021opening}.
These tools reduce the manual effort needed to construct a comprehensive leakage contract for a given CPU by automating the characterization of instruction-level leaks, whereas 
other classes of leaks can be characterized using existing largely manual approaches~\cite{hofmann2023speculation}.

However, existing contract synthesis approaches~\cite{wang2025synthesis, hsiao2024rtl2m, dinesh2025h, DBLP:conf/date/MohrG024, DBLP:conf/sp/DineshPF24} suffer from two core limitations.
First, they require access to a CPU's design at Register-Transfer Level (RTL) and, so far, have been applied only to small RISC-V cores.
This limits their applicability to complex commercial CPUs (e.g., the latest x86 and ARM cores), where subtle microarchitectural leaks are prevalent.
Second, they require a user-provided \emph{contract template} that directly defines the leaks that can be captured by all contracts in the synthesis' search space.
For instance, the template might allow capturing leaks that depend on whether ``an operand's value is 0.''
This limits the scope of these tools since they won't be able to capture leaks that are not explicitly part of the template.
E.g., a timing leak introduced by a computation simplification optimization on multiplications where the multiplication unit short-circuits whenever operands are 0 or 1 cannot be captured by a template that only allows expressing ``an operand's value is 0.''
Although some works~\cite{wang2025synthesis,DBLP:conf/date/MohrG024} suggested to manually tailor the template to account for missed leaks, this is time-consuming (as it often requires tinkering with the tool's internal implementation) and significantly reduces automation.

To address these issues, we propose a \emph{template-free} methodology to extract instruction-centric leakage contracts from black-box CPUs, which we implement in the \tool{} contract synthesizer.
Since \tool{} is black-box, it can directly work with commercial CPUs for which the RTL code is not available.
Furthermore, by being template-free, \tool{} overcomes the limitations of template-based approaches and it can synthesize instruction-centric contracts with minimal manual intervention.
Next, we overview our contributions. %

\compactpara{Contract synthesis methodology (\Cref{sec:approach})}
We propose a meth\-od\-ol\-o\-gy for automatically synthesizing instruction-centric leakage contracts based on hardware observations extracted from a black-box CPU.
For this, we adopt a counterexample-driven synthesis method that refines candidate  contracts based on observed hardware behavior. 
Our methodology consists of running the following two steps until reaching a fixed point:\looseness=-1
\begin{asparaitem}

    \item \textit{Step 1:} Given a candidate contract (initially empty) capturing the leaks observed so far, we use existing relational leakage testing tools~\cite{oleksenko2022revizor,scamv} to validate the contract against a black-box CPU.
    These tools generate test programs and run them on the underlying CPU to discover \emph{new microarchitectural leaks} not yet captured by the candidate contract.

    \item \textit{Step 2:} Newly discovered leaks are synthesized into new ISA-level contract clauses, which are 
    added to the original contract, ensuring that the candidate captures the discovered leak.
    To ensure that synthesized clauses precisely characterize the leak, our synthesis approach relies on both \emph{counterexamples} (i.e., pair of executions showcasing a leak) and \emph{positive examples} (i.e., pair of executions that are indistinguishable for a microarchitectural attacker).  

\end{asparaitem}

\looseness=-1

\compactpara{\tool{} synthesis tool (\Cref{sec:implementation})}
We implement our methodology in a tool called \tool (MicroArchitectural Leakage COntract Synthesizer). 
\tool incrementally builds instruction-centric leakage contracts that capture leaks in a given black-box CPU. 
While \tool{} targets {\tt x86} and {\tt ARM} architectures, the approach is general and can be adapted to other CPUs. %
To generate test cases and detect leaks (step 1), \tool leverages any suitable architecture-specific relational leakage testing tool; our implementation uses \revizor~\cite{oleksenko2022revizor,oleksenko2023hide} for {\tt x86} and \scamv~\cite{scamv,scamv:micro} for {\tt ARM}. Contract synthesis (step 2) is performed using the Rosette framework~\cite{torlak2013growing}.

Differently from existing contract synthesis approaches~\cite{DBLP:conf/sp/DineshPF24,dinesh2025h, DBLP:conf/date/MohrG024,hsiao2024rtl2m,wang2025synthesis}, \tool{} operates without access to the processor's RTL design.
Therefore, \tool{} can be applied to off-the-shelf commercial CPUs for which RTL is not available and which lack comprehensive vendor-supplied leakage specifications.
Furthermore, \tool{} goes beyond prior work by synthesizing fine-grained contracts instead of simply classifying instructions as safe or unsafe like~\cite{DBLP:conf/sp/DineshPF24,dinesh2025h}.
Finally, it guarantees that the synthesized contracts capture \emph{all} observed leaks (differently from \citet{DBLP:conf/date/MohrG024} whose contracts may miss leaks, compromising their usefulness for secure programming) and it does not require user-provided contract templates (unlike \citet{wang2025synthesis}).

\compactpara{Evaluation  (\Cref{sec:evaluation})}
To validate our methodology, we evaluated the accuracy of contracts generated by \tool on two fronts.

First, 
using five different contract models (ranging from the simple \textit{constant-time}~\cite{almeida2016verifying} model to ones capturing advanced optimizations like \textit{register file compression}~\cite{vicarte2021opening,ccs24} and \emph{silent stores suppression}~\cite{vicarte2021opening})  as ground truth, 
we used \tool{} to learn leakage contracts. %
Our results show that \tool is able to synthesize contracts that capture the ground-truth leaks and that these contracts have \emph{high precision}, i.e., they do not unnecessarily over-approximate  actual leaks. 
Furthermore, learned contracts are \emph{boundedly sound}, i.e., they capture \emph{all} leaks exposed during testing.
Second, we use \tool{} to learn leakage contracts directly from an {\tt x86} and an {\tt ARM} CPUs.
Our results show that \tool{} can learn leakage clauses describing well-known leaks (e.g., data-cache leaks) directly from hardware measurements. 

Finally, we compare the contracts synthesized by \tool{} with those generated by state-of-the-art \textit{template-based} white-box synthesis tools~\cite{DBLP:conf/sp/DineshPF24,dinesh2025h, DBLP:conf/date/MohrG024,hsiao2024rtl2m,wang2025synthesis} using four different targets. The results show that our \textit{template-free} approach eliminates the inflexibility of conventional methods, leading to more precise contracts than \textit{template-based} approaches.

\compactpara{Summary of contributions}
To summarize, this paper makes the following contributions: 
\begin{asparaitem}
	\item it presents a method to extract boundedly sound and precise instruc\-tion-centric leakage contracts from black-box CPUs 
    (\Cref{sec:approach});
	\item it implements this methodology in \tool, a tool to extract leakage contracts from 
    processors without using templates (\Cref{sec:implementation});
	\item it evaluates \tool on several architectures and different contract models as ground truth (\Cref{sec:evaluation}).
\end{asparaitem}

\section{Overview}\label{sec:overview}
We now present the core aspects of \tool{} with an example.

\newcommand{\attacker}{\textsc{Atk}}
\newcommand{\cpu}{\textsc{Cpu}}
\newcommand{\clause}{\mathit{cl}}
\newcommand{\candidate}{\mathit{cand}}

\subsection{Capturing Leaks with Leakage Contracts}\label{sec:overview:leakage-contracts}

To write side-channel-free programs for a given CPU and a given microarchitectural attacker, a programmer needs to know which program executions the attacker might distinguish to ensure that no secret data is leaked.\looseness=-1

The leakage contracts framework~\cite{guarnieri2021hardware} provides a way of formally characterizing microarchitectural side-channel leaks at the ISA level, thereby enabling secure programming.
In this framework, a microarchitectural attacker is modelled by mapping each (microarchitectural) execution to a \emph{hardware trace}, i.e., the observations that an attacker can make through microarchitectural side-channels.

To capture ISA-level leaks, a leakage contract augments the ISA with a specification of the observable side-channel leaks associated with a given CPU.
It maps each (architectural) program execution to a \emph{leakage trace}, i.e., a sequence of (ISA-level) observations exposing potentially leaked information.
In \tool{}, a contract is a set of clauses $\clause := \mathit{ex}\ \mathtt{IF}\ \mathit{pr}$, where $\mathit{ex}$ specifies what is added to the leakage trace and $\mathit{pr}$ is a predicate modeling when the clause is enabled, i.e., when the
observation is added to the trace.

A CPU \emph{satisfies} a given contract for a microarchitectural attacker $\attacker$~\cite{guarnieri2021hardware}, whenever the contract captures all leaks in the underlying CPU that are observable by the attacker.
That is, any two executions that result in different hardware traces, i.e., they are distinguishable by $\attacker$, must also result in different leakage traces.

\subsection{Synthesizing Leakage Contracts}\label{sec:overview:synthesis}

\begin{figure}[t] 
  \includegraphics[width=\columnwidth]{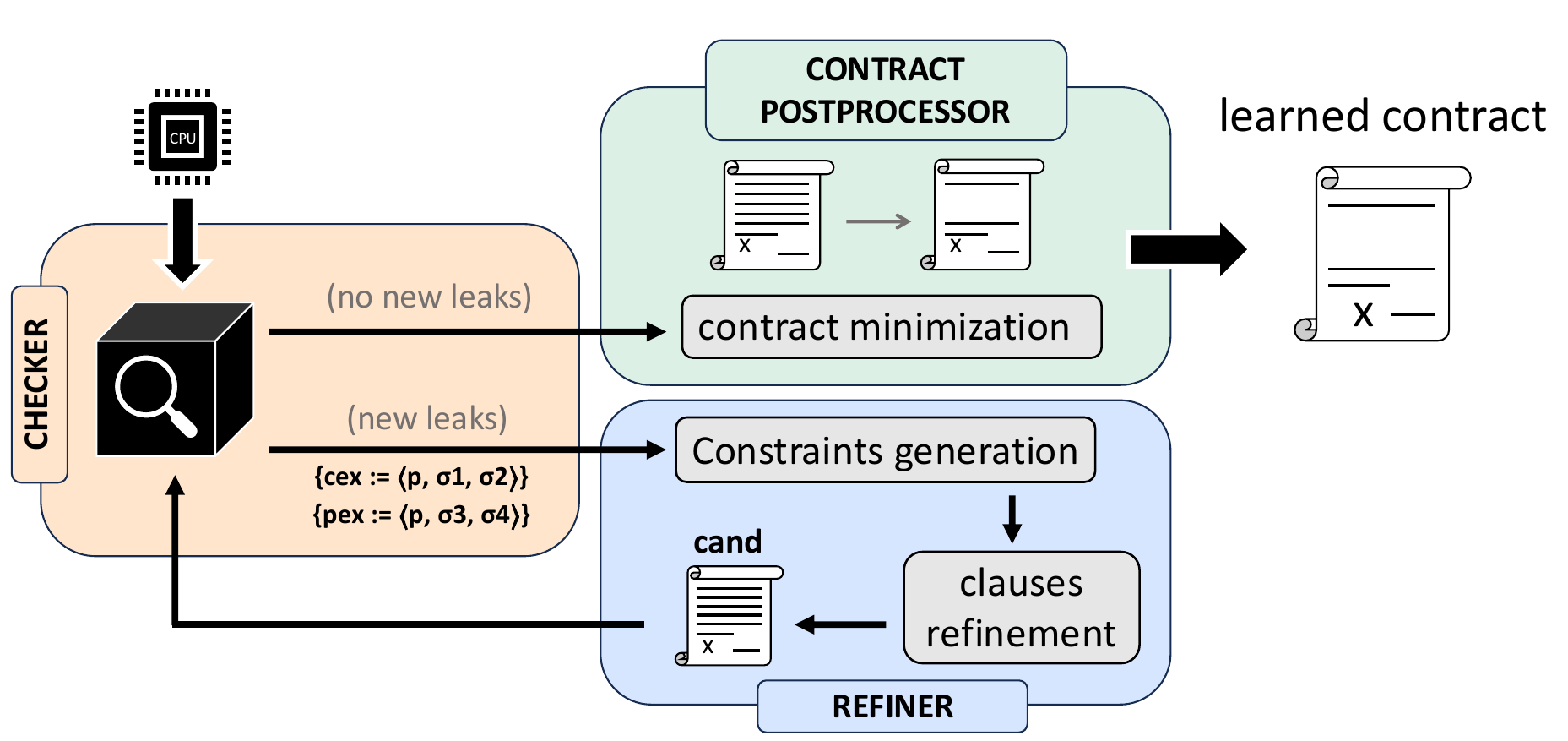} 
  \caption{\tool contract synthesis process}
  \label{fig:malcos-flow}
\end{figure}

\Cref{fig:malcos-flow} depicts \tool{}'s approach for the synthesis of leakage contracts.
The approach takes as input a black-box processor $\cpu$ and it iteratively constructs a candidate leakage contract \texttt{cand} by alternating between two phases: 
\begin{asparaenum}
    \item A \emph{leakage testing phase} (\checker  in \Cref{fig:malcos-flow}) where \tool{} attempts at finding new leaks in $\cpu$ not captured by  \texttt{cand} (capturing all leaks discovered so far).
    \item A \emph{contract refinement phase} (\refiner  in \Cref{fig:malcos-flow}) where \tool{} updates \texttt{cand} to account for a newly discovered leak.
\end{asparaenum}

When \tool{} cannot find new leaks, it simplifies the contract \texttt{cand} (\postprocessor in \Cref{fig:malcos-flow}), which is returned to the user.
Next, we provide further details on each phase.\looseness=-1

\subsubsection*{\textbf{Leakage testing phase}}
The \checker takes a candidate contract \texttt{cand} and a CPU (treated as a black-box), and tries to discover leaks not yet captured by \texttt{cand}.
Intuitively, the \checker executes programs and observes leaks w.r.t. a given attacker $\attacker$.
When it finds a new leak, it returns a \emph{counterexample} $\mathit{cex}$, which is a sequence of instructions and a pair of initial states that yield the \emph{same leakage trace} under \texttt{cand} but different hardware traces, i.e., distinguishable by $\attacker$.
It also returns \emph{positive examples} $\mathit{pex}$, i.e., test cases indistinguishable for both the contract and the attacker.

Our synthesis methodology is \checker-agnostic.
To emphasize its flexibility and generality, \tool{} uses two state-of-the-art relational leakage testers: \revizor~\cite{oleksenko2022revizor} and \scamv~\cite{scamv}.
These tools differ in methodology (\revizor{} uses fuzzing, \scamv symbolic execution), supported leakage sources (\revizor{} handles caches and timing; \scamv{} supports caches), and architecture that they target (\revizor{} targets {\tt x86}, \scamv{} targets {\tt ARM}). %

\begin{step}[A cex and a pex for RFC]\label{example:overview:checker}\label{example:overview:cpu-attacker-traces}
Consider a processor  $\cpu$
that implements a simple {\textit{register file compression}} (RFC) optimization~\cite{vicarte2021opening}.
This optimization reduces the physical size of the register file by mapping all logical registers that store the value $0$ to the same physical zero register.
As pointed out in~\cite{vicarte2021opening,ccs24}, this optimization can result in timing leaks (due to reducing the pressure on the register file).\looseness=-1
Throughout this section we consider a microarchitectural attacker $\attacker$ that can precisely observe whenever RFC happens during execution.

Starting from an empty candidate contract ($\texttt{cand} = \emptyset$), the \emph{\checker} attempts to discover a leak. %
For this, the \checker generates test cases, each one consisting of a program and a pair of initial states, executes them on the target CPU, and computes the hardware traces to detect potential leaks.
Given that  $\texttt{cand} = \emptyset$, the \checker discovers the test case $(p, \sigma_1, \sigma_2)$ as a counterexample (\texttt{cex}):
\[ 
    p := \texttt{MOV rax,rbx} \quad \sigma_1 := ( \texttt{rbx} \mapsto 0 ) \qquad \sigma_2 := ( \texttt{rbx} \mapsto 1 )
\]
Here, the program $p$ consists of an instruction assigning to register $\texttt{rax}$ the value of register $\texttt{rbx}$, where the value of $\texttt{rbx}$ is $0$ in initial state $\sigma_1$ and a value different from $0$ in initial state $\sigma_2$.
Therefore, RFC happens when executing $p$ from $\sigma_1$, but does not happen when executing $p$ from $\sigma_2$, which results in the same leakage traces w.r.t. \texttt{cand} but different hardware traces for the attacker $\attacker$.

At the same time, the \checker also discovers the test case $(p, \sigma_2, \sigma_3)$, where $\sigma_3 := ( \texttt{rbx} \mapsto 2 )$, as a positive example, since the two executions result in indistinguishable hardware \emph{and} leakage traces, i.e., the attacker $\attacker$ cannot distinguish them.
\looseness=-1
\end{step}

\subsubsection*{\textbf{Contract refinement phase}}
The \emph{\refiner} takes the counterexample \texttt{cex} (describing a newly-discovered leak) and the positive example \texttt{pex} (describing executions that should \emph{not} be distinguished), and generates a new contract clause that captures the new leak.
The \refiner discovers such a contract clause as a syntax-driven synthesis task implemented on top of the Rosette solver~\cite{torlak2013growing}.

Positive examples are essential for \emph{precision}. Without them, the \refiner may yield an over-approximate clause that exposes more information than needed (e.g., exposing \emph{all} written register values rather than only writes of value $0$).
The \refiner ensures that the generated clause distinguish \emph{as many counterexamples} as possible while distinguishing \emph{as few positive examples} as possible.
Finally, the \refiner adds the newly generated clause to the original contract to generate a new candidate contract \texttt{cand}.
This process iterates, i.e., checking and refining to discover leaks not yet captured, until no further leaks are found.

\begin{step}[A leakage contract for RFC]\label{example:overview:contract}
The \refiner analyzes the counterexample $\texttt{cex}:= \langle p, \sigma_1, \sigma_2\rangle$  and the positive example $\texttt{pex}:= \langle p, \sigma_2, \sigma_3\rangle$ of \Cref{example:overview:checker} to generate a clause capturing the leak.
It uses the Rosette solver to identify a new clause $\mathit{cl}$ of the form $\mathit{ex}\ \mathtt{IF}\ \mathit{pr}$ that distinguishes executions of $p$ from $\sigma_1$ and $\sigma_2$, while not distinguishing executions of $p$ from $\sigma_2$ and $\sigma_3$. 
This yields the following contract $\mathit{cl_1}$:
\begin{align*}
	\mathit{cl_1}
	\coloneqq\
	\setulcolor{orange}\ul{
		\textit{post-operand-value(0)}
	} 
	\texttt{ IF } &
	[
		(\greenul{\textit{operand-type(0)} \texttt{=} \texttt{reg}}) 
		\\ \texttt{ and }
		&
		(\blueul{\textit{operand-access(0)} \texttt{=} \texttt{write}}) 
		\\ \texttt{ and } 
		&
		(\redul{\textit{post-operand-value(0)} \texttt{=} \texttt{0}})
	]
\end{align*}

This contract exposes on the leakage trace \orangeul{the final value of the first operand} (index $0$) whenever \greenul{the first operand is a register} 
and it is \blueul{written} with a \redul{value $\tt 0$}.
Crucially, $\texttt{pex} := \langle p, \sigma_2, \sigma_3 \rangle$, where both states assign non-zero values to $\texttt{rbx}$, so $\attacker$ cannot distinguish them, prevents the \refiner from generating a coarser clause, yielding a \emph{precise} $\mathit{cl_1}$ that only fires when RFC actually occurs.
\end{step}

\subsubsection*{\textbf{Contract post-processing}}
Afterwards, \tool{} invokes the \postprocessor to
minimize the contract by unifying different clauses and by
reducing the number of unnecessary clauses to simplify the final (\texttt{learned}) contract, further improving its precision.
If \cpu{} has no leaks beyond RFC, $\{ \clause_1 \}$ (which exposes all registers written with a value of zero during execution) is returned as final contract.\looseness=-1

\section{Leakage Contracts}\label{sec:formalization}

This section formalizes the syntax (\Cref{sec:formalization:syntax}) and the semantics (\Cref{sec:formalization:semantics}) of contracts, as part of a language we dub \icl, and the notion of contract satisfaction~\cite{guarnieri2021hardware} (\Cref{sec:formalization:ctrsat}), which is used to detect counterexamples and positive examples.

\subsection{\langName{} Syntax}\label{sec:formalization:syntax}

\begin{figure}
  {\small\centering
  \resizebox{\columnwidth}{!}{
  \begin{tabular}{lllll}
  & & & & \\
    {\it (Bitstrings)} & ${\it bs}$ & $\coloneqq$ & ${\{0,1\}}^{n}$ \\
    {\it (Expressions)} & $\mathit{ex}$ & $\coloneqq$ &
    $\mathit{bs}$ $\mid$ 
      $\ominus \mathit{ex} \mid \mathit{ex}_1 \oplus \mathit{ex}_2 \mid \mathit{ex}[\mathit{bs}_1:\mathit{bs}_2]$ \\
  & & $\mid$ & $\mathtt{REG}(\mathit{bs}) \mid \mathtt{OPCODE} \mid \mathtt{OP\_TYPE}(\mathit{bs}) $ \\
  & & $\mid$ & $ \mathtt{OP\_ACC}(\mathit{bs}) \mid \mathtt{OP\_VAL}(\mathit{bs}) \mid $ \\
  & & $\mid$ & $ \mathtt{POST\_REG}(\mathit{bs}) \mid \mathtt{POST\_OP\_VAL}(\mathit{bs})$\\
  {\it (Predicates)} & $\mathit{pr}$ & $\coloneqq$ & \multicolumn{2}{l}{${\{0,1\}} \mid \mathit{ex}_1 = \mathit{ex}_2 \mid \mathit{ex}_1 < \mathit{ex}_2$} \\
  & & $\mid$ & \multicolumn{2}{l}{$\mathtt{NOT}\ \mathit{pr} \mid \mathit{pr}_1\ \mathtt{AND}\ \mathit{pr}_2 \mid \mathit{pr}_1\ \mathtt{OR}\ \mathit{pr}_2$} \\
  {\it (Clause)} & ${\it cl}$ & $\coloneqq$ &   \multicolumn{2}{l}{$\mathit{ex}\ \mathtt{IF}\ \mathit{pr}$}
  \\
   {\it (Contract)} & ${\it C}$ & $\coloneqq$ & \multicolumn{2}{l}{$\emptyset \mid 
   {\it C, cl}
   $} 
   \\
  \end{tabular}
  }
  }
  \caption{\langName{} syntax}
  \label{lst:contract-language}
  \end{figure}

The syntax of \langName{} is given in \Cref{lst:contract-language}.
In \langName{}, a \emph{contract} consists of a set of leakage clauses (using the terminology of~\cite{guarnieri2021hardware,oleksenko2022revizor,ccs24}), where each clause describes a piece of information that might be leaked through side channels by the underlying CPU.
Each clause $\mathit{cl}$ consists of a conditional expression $\mathit{ex}\  \mathtt{IF}\ \mathit{pr}$ where the expression $\mathit{ex}$ indicates the leaked information and the predicate $\mathit{pr}$ expresses under which conditions the leak happens.

The basic type in \langName{} is bit strings $\mathit{bs}$,
with bitvectors of length one representing booleans.
Expressions $\mathit{ex}$ are constructed by combining bitstrings $\mathit{bs}$ with unary $\ominus \mathit{ex}$ and binary $\mathit{ex}_1 \oplus \mathit{ex}_2$ operators, as well as the slice operator $\mathit{ex}[\mathit{bs}_1:\mathit{bs}_2]$, which extracts from $\mathit{ex}$ the substring between indices $\mathit{bs}_1$ and $\mathit{bs}_2$.
Additionally, expressions can refer to five pre-defined functions:
\begin{asparaitem}
\item 
$\mathtt{REG}(\mathit{bs})$ looks up into the (logical) register file and returns the value associated with the register with index $\mathit{bs}$.
\item 
$\mathtt{OPCODE}$ returns the opcode of the instruction.
\item 
$\mathtt{OP\_TYPE}(\mathit{bs})$ returns the type, which can be one of $\{\mathtt{reg},\mathtt{mem},\mathtt{imm}\}$, of the operand in position $\mathit{bs}$.
\item 
$\mathtt{OP\_ACC}(\mathit{bs})$ returns the access mode, which can be one of $\{\mathtt{r},\mathtt{w},\mathtt{rw}\}$, of the operand in position $\mathit{bs}$.
\item 
$\mathtt{OP\_VAL}(\mathit{bs})$ returns the value of the operand in position $\mathit{bs}$.  This value depends on $\mathtt{OP\_TYPE}(\mathit{bs})$ so that, if it is a register $\mathtt{reg}$, the operand value will be the value of the register, if it is memory $\mathtt{mem}$, the operand value will be the memory address it points to, and if it is an immediate $\mathtt{imm}$, the operand value will be the value of the immediate. %
\item 
$\mathtt{POST\_REG}(\mathit{bs})$ and $\mathtt{POST\_OP\_VAL}(\mathit{bs})$  returns the value of oper\-ands and registers after executing the instruction. %
\end{asparaitem}

Finally, predicates can be boolean values, equalities and inequalities between expressions, or combinations of predicates with the standard logical operators.

\subsection{\langName{} Semantics}\label{sec:formalization:semantics}

The key element of interest in the semantics of leakage contracts is the \emph{leakage trace}, i.e., the sequence of observations obtained by evaluating the clauses in the contract on \emph{every state} in the execution.
In order to collect these traces (\Cref{sec:leaktr}), \icl requires the definition of a system semantics, which captures how the underlying system works (\Cref{sec:syssem}) and of a contract semantics, which dictates how the observations that constitute a leakage trace are generated (\Cref{sec:consem}).

This separation of semantics is unlike prior work on leakage contracts~\cite{guarnieri2021hardware}, where system and contract semantics are fused in one.
This separation allows \icl to be modular in the underlying system semantics, for example, it can be instantiated with an ISA semantics such as the one of $\mu$-assembly~\cite{guarnieri2020spectector} or with a speculative semantics~\cite{xaverccs}.

\subsubsection{System Semantics}\label{sec:syssem}
The system semantics $\xrightarrow[]{}:\Sigma\times\Sigma$ maps each state 
to the next one
by executing one instruction in program $p$ (written $p \vdash \sigma \xrightarrow{} \sigma'$), where $\Sigma$ is the set of all states and $\sigma \in \Sigma$ is a system state.
The system semantics is left abstract except for these requirements:
First, the set $\Sigma_0 \subseteq \Sigma$ denotes the set of initial states.
Second, states $\sigma \in \Sigma$ contain a memory $m$ and a register assignment $r$ that, respectively, represent the state of memory and registers.
Memories $m$ map addresses (represented as integers) to values. 
Register assignments $r$ map register identifiers (represented as bitstrings) to their values.
We also assume that one of the register identifiers, denoted $\mathtt{PC}$, represents the program counter register.

\subsubsection{Contract Semantics}\label{sec:consem}
The semantics of an \langName{} contract $C$ is obtained by extending the system semantics with observations $\delta \in \mathit{Obs}$ generated according to $C$.
Formally, each \langName{} contract $C$ induces a labeled semantics $\xrightarrow[]{}_{C}:\Sigma\times\mathit{Obs}\times\Sigma$ specified by this inference rule:
\[
\footnotesize
\frac{p \vdash \sigma \xrightarrow[]{} \sigma' \quad \delta=\{ \llbracket \mathit{e} \rrbracket (\sigma,\sigma') \mid \mathit{e}\ \mathtt{IF}\ \mathit{p} \in C \wedge \llbracket \mathit{pr} \rrbracket (\sigma,\sigma') = \top \}}{p \vdash \sigma {\xrightarrow[]{\delta}_{C}} \sigma'}
\]
\noindent where $\llbracket \mathit{ex} \rrbracket (\sigma,\sigma')$ (respectively, $\llbracket \mathit{pr} \rrbracket (\sigma,\sigma')$) is the result of evaluating expression $\mathit{ex}$ (respectively, predicate $\mathit{pr}$) in states $\sigma$ and $\sigma'$. 
Note that  $\llbracket \cdot \rrbracket$ needs the next-state $\sigma'$ to evaluate $\mathtt{POST\_REG}(\mathit{bs})$ and $\mathtt{POST\_OP\_VAL}(\mathit{bs})$.
In a nutshell, the rule above labels a step of the system semantics ($p \vdash \sigma \xrightarrow[]{} \sigma'$) with the set $\delta$ of observations obtained by evaluating in the current state the expressions of all leakage clauses in $C$ whose predicates are satisfied.\looseness=-1 %

\subsubsection{Leakage Traces}\label{sec:leaktr}
A \emph{leakage trace} $\tau$ is a sequence of observations obtained by applying the contract semantics $\xrightarrow[]{}_{C}$ to an execution $p \vdash \sigma_0 \xrightarrow{} \sigma_1 \xrightarrow{} \ldots$, that is, $\tau = \delta_0 \delta_1 \dots$ where $p \vdash \sigma_0 \xrightarrow{\delta_0}_{C} \sigma_1 \xrightarrow{\delta_1}_{C} \ldots$
In the following, we write $\ctrace{C, p, \sigma}$ to denote the leakage trace (under contract $C$) associated with the maximal execution of program $p$ starting from initial state $\sigma$.
Furthermore, we say that two initial states $\sigma$, $\sigma'$ are \emph{$C$-equivalent} for a program $p$ whenever they result in the same leakage traces.
Formally: 
$
\pceq{p}{\sigma}{\sigma'} \isdef \ctrace{C, p, \sigma} = \ctrace{C, p, \sigma'}.
$

\subsection{Contract Satisfaction}\label{sec:formalization:ctrsat}
Contract satisfaction precisely characterizes under which conditions a contract captures all leaks in a CPU.

Following~\cite{guarnieri2021hardware,oleksenko2022revizor}, we first introduce the notion of hardware traces, which capture the observational power of the microarchitectural attacker. %
We represent the hardware trace as the output of the function
$\htrace{p, \sigma, \mathit{Ctx}}$,
that returns the observations made by the attacker
through microarchitectural side-channels. 
The function  $\htrace{\cdot}$ takes as inputs
the victim program $p$,
the initial state $\sigma$ processed by the victim program (i.e., the architectural state including registers and main memory),
and the microarchitectural context $\mathit{Ctx}$ in which it executes (i.e., the initial state of microarchitectural components like caches and predictors).

Informally, a CPU \emph{satisfies a leakage contract $C$ for a program $p$}~\cite{guarnieri2021hardware} if any two $C$-equivalent states also result in identical hardware traces in any context, i.e., the executions are indistinguishable by the attacker.
Formally:
\begin{align*}
\csat{p}{C} \isdef&\
\forall \sigma,\sigma',\mathit{Ctx}\ldotp
\\
&
\text{ if }
\pceq{p}{\sigma}{\sigma'}
\text{ then }
\htrace{p, \sigma, \mathit{Ctx}} = \htrace{p, \sigma', \mathit{Ctx}} 
\end{align*}
We say that a CPU \emph{satisfies a contract $C$} if the CPU satisfies it for every program.
Dually, if a CPU does not satisfy a contract $C$, there is a \textit{leakage counterexample} $\texttt{cex}:= \langle p, \sigma, \sigma'\rangle$ consisting of a program $p$ and two initial states $\sigma, \sigma'$ that produce different hardware traces (for some $\mathit{Ctx}$) but it generates identical leakage traces for $C$.

\section{Synthesizing Leakage Contracts}\label{sec:approach}
In this section, we present our methodology for automatically synthesizing instruction-centric leakage contracts. 

\subsection{Synthesis Algorithm} \label{sec:synthesis:algorithm}

Our algorithm (shown in \Cref{algo:contract-synth1})  automatically constructs a leakage contract from a CPU under test (denoted $\mathit{target}$ in this section) in a black-box manner.
That is, the \checker does not have access to the CPU design, and it only interacts with the CPU by executing \emph{test cases}, each one consisting of a program $p$ and a sequence of initial states,  to derive the corresponding hardware traces.\looseness=-1

\begin{algorithm}[t]
    \caption{\tool{} contract synthesis approach}
    \label{algo:contract-synth1}
    \small
    \SetKwProg{Proc}{Procedure}{:}{}
    \SetKwRepeat{Do}{do}{while}
    \SetKwProg{EProc}{Procedure}{()}{}

    \Proc{\textsc{ContractSynthesis}($\mathit{target},n, r$)}{
        \tcp{Phase 1 - synthesis} 
        $P \leftarrow \texttt{generateSeedPrograms}(n)$\; \label{line:synth:phase1:begin} \label{line:synth:seeds}
        $\mathit{cand} \leftarrow []$\;
        $\mathit{ctr} \leftarrow []$\;
        $\mathit{CEXs} \leftarrow []$\;
        \For{$p \in P$}{
            $\mathit{cand}, \mathit{ctr} \leftarrow \mathtt{getInitialCand}(\mathit{ctr}, \mathit{cand}, \mathit{r})$\;\label{line:synth:init-cand:begin}
            \While{ $\mathbf{true}$}{
                $\mathit{cexs}, \mathit{pexs} \leftarrow \texttt{Checker}(p, \mathit{cand}, \mathit{target})$\;\label{line:synth:ctrsatoracle}
                $\mathit{CEXs}[p] \leftarrow \mathit{CEXs}[p] \cup \mathit{cexs}$\;
                \uIf{$\mathit{cexs}\neq \emptyset$}{
                    $\mathit{cand} \leftarrow \texttt{Refiner}(p, \mathit{cand},\mathit{cexs}, \mathit{pexs})$\;\label{line:synth:refine}
                }
                \Else{
                    \textbf{break}
                }
            }
            \label{line:synth:phase1:end}
        }
        \tcp{Phase 2 - minimization} 
        $\mathit{ctr} \leftarrow \mathtt{append}(\mathit{ctr}, \mathit{cand})$\;\label{line:synth:phase2:begin}
        $\mathit{ctr} \leftarrow \mathtt{minimize}(\mathit{ctr}, \mathit{CEXs})$\;  \label{line:synth:minimization}
        \KwRet $\mathit{ctr}$\; \label{line:synth:phase2:end}
    }

\end{algorithm}

The synthesis algorithm takes three inputs---
the CPU under test $\mathit{target}$, the number of test programs $n$ used during the synthesis, and the reset parameter $\mathit{r}$---and it works in two phases. %

Phase 1 spans lines \ref{line:synth:phase1:begin}--\ref{line:synth:phase1:end}, and it synthesizes a set of
\langName{} candidate contracts that capture the leaks explored by $n$ test programs.
Phase 1 starts by computing a set $P$ of $n$ (randomly generated) test programs (line \ref{line:synth:seeds}).
For each program $p \in P$, the algorithm first computes the starting initial candidate contract $\mathit{cand}$ (line \ref{line:synth:init-cand:begin}).
This candidate can be the empty contract $\emptyset$, or it can be derived from the contracts generated from prior programs.
{Every $r$ iterations, the current set of candidates $\mathit{cand}$ is added to the accumulated contract $\mathit{ctr}$, and the search is restarted from the empty candidate, thereby allowing \tool{} to recover from mistakes caused by  low-quality examples.} %
Next, the algorithm checks whether there are leaks in the underlying CPU under test  that are not yet captured  by the candidate contract $\mathit{cand}$ (line \ref{line:synth:ctrsatoracle}, described in \Cref{sec:synthesis:ctrsatoracle}).
For this, the algorithm relies on the \checker function, which returns a set of counterexamples $\mathit{cexs}$ and a set of positive examples $\mathit{pexs}$ for the program $p$.
The former are leaks not yet captured by the contract $\mathit{cand}$, whereas the latter represent executions that are indistinguishable for the microarchitectural attacker (which are used to improve the precision of the synthesized contract).
If \checker finds at least one counterexample, the algorithm refines the current candidate contract $\mathit{cand}$ to account for the newly discovered leaks using the \texttt{Refiner} function (line \ref{line:synth:refine}; described in~\Cref{sec:synthesis:refinement}). 
If \checker cannot find any further counterexamples, the algorithm terminates the testing of the current program,
and moves to the next program in $P$.\looseness=-1

Phase 2 spans lines~\ref{line:synth:phase2:begin}--\ref{line:synth:phase2:end}, and {it first consolidates the remaining candidates $\mathit{cand}$ into the set of $\mathit{ctr}$ contracts, then it } 
derives the final leakage contract for the target CPU by minimizing $\mathit{ctr}$ using 
the \texttt{minimize} function (line \ref{line:synth:minimization}; described in \Cref{sec:synthesis:minimization}). 
The minimization step allows us to remove unnecessary leakage clauses from the contract, while preserving soundness w.r.t. the explored counterexamples.
That is, the final leakage contract still distinguishes all counterexamples in $\mathit{CEXs}$, and it is more precise (i.e., it distinguishes fewer pairs of states with the same hardware trace) than the non-minimized contract. 
Finally, this phase returns the final leakage contract $\mathit{ctr}$ for the target CPU to the user.

\subsection{Choosing the Initial Candidate Contract}\label{sec:synthesis:initial} 
To select the initial candidate contract when testing a new program ($\mathtt{getInitialCand}$ on line \ref{line:synth:init-cand:begin} of \Cref{algo:contract-synth1}), \tool combines a strategy of 
\textit{integration} and \textit{resetting} based on a parameter $r$:
 the initial contract $\mathit{cand}$ for the $i$-th tested program is obtained by integrating the contracts synthesized for the last $i\ \texttt{mod}\ r$ programs.\looseness=-1

This strategy allows the new candidate to account for previously found leaks--through the integration of previous contracts--and, therefore, reduces the time needed to find a candidate that covers all program leaks. 
At the same time, by 
resetting the candidate every $r$ programs, it allows \tool{} to recover from mistakes.
For instance, it allows resetting the contract when the \checker has not provided good enough positive examples for an individual program, which may lead the synthesis to over-approximate the leakage.

\subsection{Finding Leakage Counterexamples}\label{sec:synthesis:ctrsatoracle}

When learning a new contract, our synthesis algorithm relies on a \checker that interacts with the CPU to determine whether it satisfies the current candidate contract.
The \checker function takes as input the current program $p$, the contract candidate $\mathit{cand}$ and the target CPU $\mathit{target}$, and it checks whether $\mathit{target}$ satisfies the contract $\mathit{cand}$ w.r.t. the contract satisfaction notion from \Cref{sec:formalization:ctrsat}.
Concretely, the \checker generates a \emph{test case} for the program $p$ (i.e., a set of initial states for $p$), derives the corresponding \emph{hardware traces} and  \emph{contract traces} from $\mathit{target}$ and $\mathit{cand}$ respectively, and outputs a set of counterexamples and a set of positive examples. 

A \emph{counterexample} consists of two $\mathit{cand}$-equivalent states $\sigma, \sigma'$  with different hardware traces, i.e., a microarchitectural attacker can distinguish the executions of $p$ starting from $\sigma$ and from $\sigma'$ through a microarchitectural side-channel.
That is, each counterexample represents a leak not captured by the contract $\mathit{cand}$.\looseness=-1

In contrast, a \emph{positive example} consists of two $\mathit{cand}$-equiv\-a\-lent states $\sigma, \sigma'$ that additionally have the same hardware traces, i.e., two states that are indistinguishable for a microarchitectural attacker.
Each positive example represents a pair of executions that the final contract should not distinguish.
That is, if the final contract distinguishes two executions forming a positive example, the contract is an over-approximation of the desired contract (i.e., it distinguishes more executions than needed).

We remark that our synthesis algorithm is not tied to a specific implementation of \checker. 
Rather, it can work on top of any approach for checking contract satisfaction.
As we discuss in \Cref{sec:implementation}, our implementation \tool{} instantiates \checker using two state-of-the-art black-box testing tools for contract satisfaction: Revizor~\cite{oleksenko2022revizor} for synthesis over the x86 architecture, and Scam-V~\cite{scamv,scamv:micro} for synthesis over ARM architectures.

\subsection{Contract Refinement}\label{sec:synthesis:refinement}

The \texttt{Refiner} function takes as input a candidate $\mathit{cand}$, the test program $p$, a set of contract counterexamples $\mathit{cexs}$ and as set of positive examples $\mathit{pexs}$ for $p$ and $\mathit{cand}$.

In a nutshell, the function refines the current candidate contract $\mathit{cand}$ to account for newly discovered leaks represented by the counterexamples in $\mathit{cexs}$.
This  requires finding an \icl contract $\mathit{cand}'$ such that (a) pairs of executions that are $\mathit{cand}$-distinguishable are also $\mathit{cand}'$-distinguishable (i.e., the refined contract is not ``forgetting'' leaks), 
and (b) $\mathit{cand}'$ distinguishes {at least one} counterexample in $\mathit{cexs}$ (i.e., the refined contract indeed captures {at least one} newly discovered leak).\looseness=-1

For this, we cast contract refinement as an optimization syntax-guided synthesis problem. %
In particular, we refine the contract $\mathit{cand}$ by synthesizing at most $m$ additional leakage clauses $\mathit{cl}_1, \ldots, \mathit{cl}_m$, where $m$ is a parameter of the refinement procedure.
Each clause $\mathit{cl}_i$ has to satisfy the following two hard-constraints, where $?$ denotes a hole to be filled by the synthesis solver according to \langName{}'s grammar:
\begin{align}
    \mathit{cl}_i = (?\ \mathtt{IF}\ \mathtt{REG}[\mathbf{PC}] =\ ?\ \wedge\ ?) \label{harcon-1}
    \\
    \bigvee_{(p,\sigma,\sigma') \in \mathit{cexs}} \ctrace{\mathit{cl}_i,p,\sigma} \neq \ctrace{\mathit{cl}_i,p,\sigma'} \label{harcon-2}
\end{align}
Hard-constraint \ref{harcon-1} ensures that $\mathit{cl}_i$ is an instruction-centric clause, which targets the leak associated with a single instruction in $p$ (captured by the $\mathtt{REG}[\mathbf{PC}] =\ ?$ part of the predicate, allowing us to capture fine-grained leaks that may depend on a specific instruction~context).
In contrast, hard-constraint \ref{harcon-2} ensures that $\mathit{cl}_i$ captures the leaks of at least one of the counterexamples $\mathit{cexs}$ by requiring that $\mathit{cl}_i$ produces distinguishable leakage observations for at least one pair of counterexample states.

Additionally, to ensure the precision of $\mathit{cl}_i$, we add the following soft-constraint, which the solver should maximize:
{
\begin{align}
\mathit{max} \bigg(\!\!\! 
\begin{array}{l}
\big| \{ (p,\sigma, \sigma')\! \in\! \mathit{cexs} \mid \mathtt{CTR}(\mathit{cl}_i, p, \sigma)\!\neq\! \mathtt{CTR}(\mathit{cl}_i, p, \sigma') \} \big| \\
\; +\ \big| \{ (p,\sigma, \sigma')\! \in\! \mathit{pexs} \mid  \mathtt{CTR}(\mathit{cl}_i, p, \sigma)\!=\! \mathtt{CTR}(\mathit{cl}_i, p, \sigma') \} \big|\!\!\!\!
\end{array}
\bigg)
\label{softcon}
\end{align}
}
The soft-constraint \ref{softcon} ensures that 
$\mathit{cl}_i$
distinguishes most counterexamples, i.e., the one capturing more leaks.
It also ensures that 
$\mathit{cl}_i$ will be the ``most precise'' clause that satisfies constraints \ref{harcon-1} and \ref{harcon-2}, i.e., the one that distinguishes as few positive examples as possible.

The synthesis solver outputs 
 a set of instruction-centric clauses $\mathit{cl}_i$ that are specific for the program $p$.
 These clauses are of the form $\mathit{ex}\ \mathtt{IF}\ \mathtt{REG}[\mathbf{PC}] = \mathit{val} \wedge \mathit{pr}$ where $\mathit{val}$ is the address of one of $p$'s instructions. 
 That is, they refer explicitly to the context of the program $p$.
Then, the \refiner generalizes these clauses, allowing them to be applied across all programs.
Concretely, it 
replaces each clause 
$\mathit{ex}\ \mathtt{IF}\ \mathtt{REG}[\mathbf{PC}] = \mathit{val} \wedge \mathit{pr}$
with an equivalent clause
$
  \mathit{ex}\ \mathtt{IF}\ \mathit{type}_{p,\mathbf{PC}=\mathit{val}} \wedge \mathit{pr}
  $
where $\mathit{type}_{p,\mathbf{PC}=\mathit{val}}$ is a meta-level predicate matching the instruction type of the corresponding instruction at program counter $\mathit{val}$ in $p$.
Thus, $\mathit{type}_{p,\mathbf{PC}=\mathit{val}}$ is defined in \icl as the conjunction of predicates capturing the instruction type characteristics at the specific program counter $\mathit{val}$ (namely $\mathtt{OP\_TYPE}(\mathit{bs})$, $\mathtt{OP\_ACC}(\mathit{bs})$, and $\mathtt{OPCODE}(\mathit{bs})$).
That is, if the clause $\mathit{cl}$ is associated with program counter $\mathit{addr}$ pointing to instruction $i$, the corresponding generalized clause is obtained by replacing the sub-predicate $\mathtt{REG}[\mathbf{PC}] = \mathit{addr}$ with a predicate that is satisfied whenever the program counter points to an instruction of the same type as $i$ (i.e., same op-code and operand types).

{Finally, the \refiner outputs the new candidate contract $\mathit{cand}'$ as the union of $\mathit{cand}$ and the new clauses $\mathit{cl}_1, \ldots, \mathit{cl}_m$.}
This refinement satisfies  points (a) and (b) above  (the former follows from $\mathit{cand}' = \mathit{cand} \cup \{ \mathit{cl}_1, \ldots, \mathit{cl}_m\}$ and the latter from the second constraint).

\subsection{Minimization}\label{sec:synthesis:minimization}

Phase 2 of our algorithm post-processes the \langName{} contract to obtain a more precise one.
Thus, the $\mathtt{minimize}$ function removes unnecessary clauses from the  contract candidate $\mathit{cand}$.
For this, $\mathit{cand}$'s clauses are sorted in terms of their precision (from higher to lower) over the test cases explored during Phase 1, where more precise clauses distinguish less attacker-equivalent pairs of executions.
Then, the clauses that do not contribute to distinguishing the collected counterexamples are iteratively removed, i.e., a clause is removed if all counterexamples are still distinguishable by a contract containing only the remaining clauses.
This step allows us to remove ``useless'' clauses that do not capture actual leaks and only reduce precision.
This minimization step significantly contributes to improving the precision of the final contract (see \Cref{sec:eval:rq3}).

\section{Implementation}\label{sec:implementation}

This section presents our contract synthesis tool.
To show the generality of our approach, we implemented two backends to support  \texttt{x86} and \texttt{ARM} architectures. 
We refer to \tool{} with each backend as \toolx{} and \toolarm{}.
Both follow the workflow outlined in \Cref{algo:contract-synth1}, but they differ in: 
the contract satisfaction tool (\Cref{sec:implementation:oracle}), 
the way contracts are refined (\Cref{sec:implementation:refine}), and contract post-processing (\Cref{sec:implementation:post-processing}).
We remark that the differences between the two backends ultimately stem from the different contract satisfaction tools used.

\subsection{Code Examples and Contract Satisfaction}\label{sec:implementation:oracle}

The \checker part of \tool must generate test programs (function $\mathtt{generateSeedPrograms}$ in \Cref{algo:contract-synth1}) and extract counter- and positive examples for a given contract candidate.
\tool implements the \checker with tools for relational contract testing: \revizor~\cite{oleksenko2022revizor} for \texttt{x86} and \scamv~\cite{scamv} for \texttt{ARM}.
Both tools take as input a contract and try to discover contract violations (i.e., counterexamples) in the CPU under test by generating random programs and inputs, executing them on the actual CPU and at contract-level, and comparing collected traces.
Off-the-shelf, both tools only output counterexamples, so we modified them to also extract positive examples.
Each call to the $\mathtt{Checker}$ produces a single counterexample in \toolx but yields multiple counterexamples in \toolarm.

\subsubsection*{\toolx{}:}
For \texttt{x86}, we rely on \revizor, a black-box hardware fuzzer 
that uses differential testing to compare pairs of hardware traces with pairs of contract traces to detect contract violations, and we extend it to support \icl contracts. 

Given that \revizor uses randomly-generated programs and inputs during the testing process, we use two optimizations implemented in \revizor---program minimization and data minimization---to tame randomness and improve the precision of the synthesized contracts.
\emph{Program minimization} minimizes test programs that result in violations.
For this, whenever \revizor finds a counterexample for a program $p$, \revizor iteratively removes instructions from $p$ so long as it can still find a counterexample in the smaller program $p'$.
\emph{Data minimization}, instead, identifies which parts of the initial states lead to a violation.
For this, whenever \revizor finds two architectural states $\sigma$ and $\sigma'$ causing a violation, it iteratively modifies $\sigma'$ by copying part of the state from $\sigma$ (by copying registers and memory at a byte-granularity) until the violation persists.
We remark that \toolx{} applies these minimization steps automatically whenever the \checker finds a counterexample, and the minimized counterexample is then used for synthesis. %

\subsubsection*{\toolarm{}:}
For \texttt{ARM}, we use \scamv{} as \checker, which combines techniques from program verification and fuzzing to perform relational testing to validate the candidate contracts.
For a generated test program $p$, \scamv{} uses \emph{symbolic execution} to synthesize a relation that identifies which states are observationally equivalent according to the contract being validated.
Next, \scamv generates an instance of this relation in terms of two input states.
Finally, similar to \revizor{}, \scamv{} runs the generated program with two inputs on hardware to collect hardware traces.

\subsection{Contract Refinement}\label{sec:implementation:refine}
To refine the candidate contract (function $\mathtt{Refiner}$ in \Cref{algo:contract-synth1}), we implement the instruction-centric clause synthesis process described in \Cref{sec:synthesis:refinement} using Rosette~\cite{torlak2013growing}, a solver-aided language extending Racket.
We first formalize \langName{}'s syntax and semantics in Rosette, then collect all state sequences from counterexamples $\mathit{cexs}$ and positive examples $\mathit{pexs}$ to instantiate the synthesis constraints \ref{harcon-1}--\ref{softcon}.
Rosette then uses the Z3 SMT solver~\cite{de2008z3} to synthesize \langName{} clauses that satisfy these constraints.\looseness=-1

The main difference between the \toolx{} and \toolarm{} toolchains is that
\toolx{} synthesizes one leakage clause per instruction in the program under test $p$ during refinement, while \toolarm{} synthesizes a single clause for the entire program.
This difference is due to the program minimization step in \toolx{} (\Cref{sec:implementation:oracle}), which pinpoints \emph{all} instructions involved in the leak, enabling clause synthesis for each instruction.
In contrast, \toolarm{} does not minimize the test program, which may include irrelevant instructions.
By synthesizing a single clause, we allow the solver to identify leak sources without unnecessary approximations from clauses tied to unrelated instructions.

\subsection{Contract Post-Processing}\label{sec:implementation:post-processing}

Both \toolx{} and \toolarm{} minimize the final \langName{} contract into a more precise one using the minimization step from \Cref{sec:synthesis:minimization}.
Thus, both toolchains sort clauses by the number of positive examples they (unnecessarily) distinguish, ordering from those that distinguish less (i.e., more precise clauses) to those that distinguish more. This is evaluated using the notion of C-equivalence (\Cref{sec:leaktr}), where we say a clause $\mathit{cl_{2}}$ is at least as precise as $\mathit{cl_{1}}$ if $\forall (p, \sigma, \sigma')\ldotp \pceqgen{p}{\sigma}{\sigma'}{\mathit{cl_{2}}} \text{ then } \pceqgen{p}{\sigma}{\sigma'}{\mathit{cl_{1}}}$.
This implication allows us to filter out redundant or less precise clauses. The key difference between the toolchains resides on the domain in which this condition is verified, which ultimately stems from the different checkers used in each case. \toolarm uses an SMT solver to formally verify the formula above (i.e., the universal quantification ranges over all possible pairs of states in the program),
whereas \toolx{} relies on testing the condition on the set of known counterexamples (\mbox{$\forall (p, \sigma, \sigma') \in \mathit{CEXs}$}).

\section{Evaluation}\label{sec:evaluation}

In our evaluation, we address the following questions:

\begin{enumerate}
\renewcommand{\labelenumi}{\textbf{RQ\arabic{enumi}:}}
    \item How good are \tool{}' contracts? %
    \item What is the impact of contract minimization? %
    \item Can \tool{} synthesize contracts from hardware? %
    \item How do \tool{}-learned contracts compare to \textit{template-based} approaches? %
\end{enumerate}
We begin by introducing the metrics used to evaluate contract quality in \textbf{RQ1}, \textbf{RQ2} and \textbf{RQ4} (\Cref{sec:eval:metrics}),
then we describe the target leakage contracts used as ground truth in \textbf{RQ1}---\textbf{RQ2} (\Cref{sec:eval:contracts}).
Finally, we answer \textbf{RQ1}--\textbf{RQ4} in \Cref{sec:eval:rq2,sec:eval:rq3,sec:eval:rq4,sec:eval:rq5} respectively.

\subsection{Metrics}\label{sec:eval:metrics}

Leakage contracts act as binary classifiers over pairs of executions, i.e., a contract indicates whether an observer can distinguish two executions. 
To evaluate the quality of the learned contracts, we use two standard metrics for binary classification: {precision} and {soundness}.
These metrics are defined w.r.t. a \emph{validation set} $V$ consisting of test cases $(p,\sigma,\sigma')$ where $p$ is a program and $\sigma, \sigma'$ are initial states.

\emph{Precision}  measures how precisely the learned contract reflects leaks in the target system.
Following~\citet{DBLP:conf/date/MohrG024} and \citet{wang2025synthesis}, precision is defined as: 
$\frac{TP}{TP + FP},$
where $TP$ is the number of true positives, i.e., test cases in $V$ that are distinguishable using both contract and hardware traces, 
and
$FP$ is the number of false positives, i.e., test cases in $V$ that are distinguishable only by the contract traces but not by the hardware traces.

\emph{Soundness} measures the correctness of the learned contract.
We define it as $\frac{TP}{TP + FN}$,
where $FN$ is the number of false negatives, i.e., test cases in $V$ that are distinguishable by hardware traces but not by the contract traces.

\subsection{Contract Models}\label{sec:eval:contracts}
To answer \textbf{RQ1}---\textbf{RQ2}, we use \tool{} to synthesize leakage contracts against a set of fixed ground-truth contract models, i.e., \tool{} uses these models as the \emph{target} to obtain hardware traces.\footnote{
\toolx{} and \toolarm{} have been extended to support a leakage contract $C$ as synthesis target.
For this, the hardware trace associated with program $p$ and state $\sigma$ is computed by simulating $p$ using an ISA simulator and, for each instruction, all clauses in $C$ are evaluated w.r.t. the current architectural state to compute the corresponding leakage observations.
Note that, computing the hardware trace when using a given contract as synthesis target involves no hardware measurements, and the hardware trace consists only of the leakage observations computed by the ISA simulator.
}
These models, described next, capture representative microarchitectural behaviors and distinguish relevant leakage scenarios.
\begin{asparaitem}
\item \textbf{Constant-time (ct)} 
models the constant-time observer commonly used when reasoning about side channels in cryptographic algorithms~\cite{almeida2016verifying}. 
\textbf{ct} exposes the value of the program counter and the addresses of load and store operations throughout the execution.

\item \textbf{Tag Index (TagIdx)}
differs from \textbf{ct} in that it exposes only the tag and set index of  memory accesses (instead of the whole memory address like in \textbf{ct}). 
\textbf{TagIdx} is often used to reason about the security of cryptographic implementations against  cache-based side-channel attacks~\cite{doychev2015cacheaudit}.

\item \textbf{Register file compression (RFC)}
models the leaks induced by the compression optimization described in \Cref{sec:overview}.

\item \textbf{Silent Store (SilStore)}
differs from \textbf{RFC} in that it produces a leak when a value 0 is written to memory (instead of a register like \textbf{RFC}).
\textbf{SilStore} models leaks from store operations that do not alter memory contents~\cite{vicarte2021opening}, and its effects have been observed on several Intel CPUs~\cite{downs2020intel}.

\item \textbf{Multiplication simplification (mul)}
models leaks caused by  computation simplification over multiplications~\cite{vicarte2021opening}.
Specifically, \textbf{mul} exposes if a multiplication's operands are  0 or 1.

\end{asparaitem}

\begin{table}[]
    \small
    \centering
    \resizebox{\columnwidth}{!}{%
    \begin{tabular}{c|c|c|c|c|c|c}
    \specialrule{1.25pt}{0pt}{0pt}
    \multirow{3}{*}{Contract} & 
    \multicolumn{3}{c|}{\makecell[c]{\textbf{\toolx} \\ \textbf{with Ctr. minimization}}} & 
    \multicolumn{3}{c}{\makecell[c]{\textbf{\toolarm} \\ \textbf{with Ctr.minimization}}} \\ \cline{2-7}
                              & \begin{tabular}[c]{@{}c@{}}avg. \\P/S\end{tabular} 
                              & \begin{tabular}[c]{@{}c@{}}avg. \\Time\end{tabular} 
                              & \begin{tabular}[c]{@{}c@{}}avg. \\ \# Clauses\end{tabular} 
                              & \multicolumn{1}{c|}{P/S} 
                              & \multicolumn{1}{c|}{Time} 
                              & \begin{tabular}[c]{@{}c@{}}\# Clauses\end{tabular} \\ \hline \hline
    \textbf{ct}                        & \multicolumn{1}{r|}{1.0/1.0}    & \multicolumn{1}{r|}{11h 57m} & 6 & \multicolumn{1}{r|}{1.0/1.0}  
    & \multicolumn{1}{r|}{5h 40m}
    & 2 \\ \hline
    \textbf{TagIdx}                    & \multicolumn{1}{r|}{0.79/1.0}   & \multicolumn{1}{r|}{18h 57m} & 14.4 & \multicolumn{1}{r|}{1.0/1.0}  
    & \multicolumn{1}{r|}{13h 42m}
    & 2 \\ \hline
    \textbf{RFC}                       & \multicolumn{1}{r|}{1.0/1.0}    & \multicolumn{1}{r|}{10h 32m} & 9.8 & \multicolumn{1}{r|}{1.0/1.0}  
    & \multicolumn{1}{r|}{12h 51m}
    & 6 \\ \hline
    \textbf{mul}                       & \multicolumn{1}{r|}{1.0/1.0}    & \multicolumn{1}{r|}{24h 21m} & 15.7 & \multicolumn{1}{r|}{1.0/1.0}  
    & \multicolumn{1}{r|}{22h 34m} 
    & 4  \\ \hline
    \textbf{SilStore}                  & \multicolumn{1}{r|}{1.0/1.0}    & \multicolumn{1}{r|}{1h 23m}  & 1 & \multicolumn{1}{r|}{1.0/1.0}  
    & \multicolumn{1}{r|}{5h 29m} 
    & 1 \\ 
    \specialrule{1.25pt}{0pt}{0pt}
    \end{tabular}
    }
    \caption{\tool{} results: Precision (P), Soundness (S), and Number of Clauses (\# Clauses).}
    \label{tab:arch-level-results-both}
    \vspace{-1em}
\end{table}

\subsection{\textbf{RQ1:} How Good Are \tool' Contracts?}\label{sec:eval:rq2}
\textbf{RQ1} evaluates the contracts generated by \Cref{algo:contract-synth1}.

\compactpara{Experimental setup}\label{sec:eval:rq1:setup}
We use \toolx and \toolarm to synthesize 
contracts against the ground-truth models from~\Cref{sec:eval:contracts}.

For \toolx, for each target contract, we randomly generate 10000 programs and 100 random states per program, resetting the contract after every 500 programs (i.e., $R = 500$), and we use \toolx to synthesize the contracts (using 100 positive examples during synthesis). %
We repeat the experiment outlined above 10 times, with different initial randomness seeds.
We then measure \tool runtime and the precision/soundness of the learned contracts against a validation set of 50000 new random programs and 100 new random states per program (i.e., $100^2$ pairs of traces), and report the average of both metrics.

For \toolarm{}, for each target contract, we randomly generate 100 programs,
with 20 positive examples 
(cached at line \ref{line:synth:ctrsatoracle} in \Cref{algo:contract-synth1} and reused among the while loop iterations of the program under test)
and 10 counterexamples (updated for each call to the \refiner, line \ref{line:synth:refine} in \Cref{algo:contract-synth1}),
resetting the contract after every program (i.e., $R = 1$). %
We then measure \tool runtime and the precision/soundness of each synthesized contract against a validation set consisting of 100 new programs, each with 100 pairs of states (50 target-distinguishable pairs and 50 target-indistinguishable pairs).

For both \toolx{} and \toolarm{}, we report the number of clauses in the synthesized contracts, i.e., the final number of clauses after the postprocessing phase in \Cref{algo:contract-synth1}.

\compactpara{Results}
\Cref{tab:arch-level-results-both} summarizes our results. 
 Both tool\-chains successfully learn contracts that capture all target leaks, i.e., soundness is always 100\%.
Both tools resolved the over-approximations during the minimization process, achieving 100\% precision in almost all cases (see \Cref{tab:contract-min-results-both} for comparison).

We manually inspected the case where \toolx could not achieve 100\% precision (\textbf{TagIdx}, 79\%), and we observed conservative over-ap\-prox\-i\-ma\-tions that justify not reaching 100\% precision,
confirming that the lower precision is due to \toolx exposing more address bits than are actually needed in the final contract.
These over-approximations results from low-quality examples, whose impact can be mitigated by reducing the reset parameter or increasing the number of positive examples as shown in \Cref{sec:x86_best_config:pex} and \Cref{sec:x86_best_config:resetting}.
For example, \toolx with more frequent resetting ($R=100$) achieves a precision of 96\% for \textbf{TagIdx}. %

We also manually inspected all synthesized contracts that achieved 100\% precision  (for \toolx{} and \toolarm{}) and confirmed their equivalence to the ground truth. 
In particular, for \toolarm{}, we leveraged \scamv{}'s SMT-backend to verify that, indeed, synthesized contracts and target ones are equivalent.

\begin{table}[]
    \small
    \centering
    \resizebox{\columnwidth}{!}{%
    \begin{tabular}{c|c|c|c|c|c|c}
    \specialrule{1.25pt}{0pt}{0pt}
    \multirow{3}{*}{Contract} & 
    \multicolumn{3}{c|}{\makecell[c]{\textbf{\toolx} \\ \textbf{without Ctr. minimization}}} & 
    \multicolumn{3}{c}{\makecell[c]{\textbf{\toolarm} \\ \textbf{without Ctr.minimization}}} \\ \cline{2-7}
                              & \begin{tabular}[c]{@{}c@{}}avg. \\P/S\end{tabular} 
                              & \begin{tabular}[c]{@{}c@{}}avg. \\Time\end{tabular}  
                              &\begin{tabular}[c]{@{}c@{}}avg. \\ \# Clauses\end{tabular}
                              & \multicolumn{1}{c|}{P/S} 
                              & \multicolumn{1}{c|}{Time} 
                              & \begin{tabular}[c]{@{}c@{}}\# Clauses\end{tabular} \\ \hline \hline
    \textbf{ct}                        & \multicolumn{1}{r|}{0.70/1.0}    & \multicolumn{1}{r|}{10h 31m} & 288.2 & \multicolumn{1}{r|}{0.51/1.0}  & \multicolumn{1}{r|}{5h 21m}
    & 24 \\ \hline
    \textbf{TagIdx}                    & \multicolumn{1}{r|}{0.14/1.0}   & \multicolumn{1}{r|}{14h 14m}  & 388.9 & \multicolumn{1}{r|}{0.50/1.0}  & \multicolumn{1}{r|}{12h 44m}
    & 204 \\ \hline
    \textbf{RFC}                       & \multicolumn{1}{r|}{0.30/1.0}   & \multicolumn{1}{r|}{8h 40m} & 368.8 & \multicolumn{1}{r|}{0.50/1.0}  & \multicolumn{1}{r|}{11h 36m}
    & 271 \\ \hline
    \textbf{mul}                       & \multicolumn{1}{r|}{0.20/1.0}   & \multicolumn{1}{r|}{14h 58m} & 255.5 & \multicolumn{1}{r|}{0.50/1.0} & \multicolumn{1}{r|}{18h 20m}
    & 404  \\ \hline
    \textbf{SilStore}                  & \multicolumn{1}{r|}{0.13/1.0}   & \multicolumn{1}{r|}{1h 19m} & 32.2 & \multicolumn{1}{r|}{0.58/1.0} & \multicolumn{1}{r|}{5h 9m}
    & 86 \\ 
    \specialrule{1.25pt}{0pt}{0pt}
    \end{tabular}
    }
    \caption{Precision (P), Soundness (S), and Number of Clauses ($\#$ Clauses) without contract minimization.}
    \label{tab:contract-min-results-both}
\end{table}

\subsection{RQ2: What is the Impact of Contract Minimization?}\label{sec:eval:rq3}
\textbf{RQ2} focuses on evaluating the impact of contract minimization (line 16 in \Cref{algo:contract-synth1}) on the quality of the synthesized contracts.

\compactpara{Experimental setup}
To assess the impact of contract minimization, we repeat the experiments of \Cref{sec:eval:rq2} while disabling the minimization phase for \toolx{} and \toolarm{}.
We measure the average \tool{} runtime, precision/soundness of the synthesized contracts against the same validation sets as in \Cref{sec:eval:rq2}, and the number of clauses in the synthesized contracts (without minimization).\looseness=-1

\compactpara{Results}
\Cref{tab:contract-min-results-both} reports the results of our experiment.
Comparing the results in \Cref{tab:contract-min-results-both} with those in \Cref{tab:arch-level-results-both} (where the synthesized contracts have been minimized) shows that, without minimization, \tool{} synthesizes contracts that are significanlty less precise, even though soundness stays at 100\%.
That is, minimization indeed is useful and improves the precision of synthesized contracts.

These results can be understood more clearly when comparing the number of clauses in the synthesized contracts with and without minimization (cf. columns \textit{\# Clauses} in \Cref{tab:arch-level-results-both,tab:contract-min-results-both}). 
Without minimization, the final contracts contain many more clauses that might unnecessarily introduce over-approximations (thereby resulting in less precise contracts) without contributing to  soundness.

Finally, as expected, removing the minimization step reduces the total execution time (since \tool{}  can directly terminate at line 15 of \Cref{algo:contract-synth1}). %
This impact is more noticeable in more complex contracts, such as  \textbf{TagIdx} and \textbf{mul}, whereas the impact is very limited or absent for  simpler contracts like \textbf{ct} and \textbf{SilStore}.\looseness=-1

\subsection{\textbf{RQ3:}  Can \tool{} Synthesize Leakage Contracts From Actual Hardware?}\label{sec:eval:rq4}

\textbf{RQ3} focuses on whether \tool{} can be used to learn contracts from x86 (\Cref{sec:eval:rq3:x86}) and ARM (\Cref{sec:eval:rq3:arm}) CPUs.

\subsubsection{x86}\label{sec:eval:rq3:x86}
Using \toolx, we synthesize contracts for 13 subsets of the x86 ISA. 
Altogether, these subsets result in a complete base x86-64 user-level instruction set, excluding the instructions currently not compatible with \revizor.
For these subsets, we re-use \revizor{}'s initial configuration from~\cite{oleksenko2023hide} where (a) all subsets include the same 8 base arithmetic/logic instructions (including their versions with memory operands), needed by \revizor to work properly, and (b) each subset extends the base instructions with unique subset-specific instructions. See \Cref{sec:x86-subsets} for a complete list of the 13 subsets and their instructions.

\compactpara{Experimental setup}
We target four different Intel CPUs: 
i5-6500  (Skylake), 
i5-1335U (Raptor Lake), 
Ultra5-225U (Arrow Lake), 
and Ultra7-258V (Lunar Lake).
For each ISA subset, we use \toolx{} to synthesize the contract capturing the leaks associated with the subset's instructions.
Each subset was tested with 100000 randomly generated programs, and each program executed with 100 inputs,
restarting from an empty candidate every 500 programs (i.e., $R = 500$), and we used up to 100 positive examples for synthesis.
To account for speculatively executed instructions, we configure \revizor to explore at ISA-level also mispredicted branches\footnote{For this, we enabled the \texttt{cond} execution clause in Revizor, namely, using the always-mispredict speculative semantics~\cite{guarnieri2020spectector} as the system semantics $\to$ for \tool{}  (\Cref{sec:formalization:semantics}).} (to account for branch speculation) and to disable store-bypass speculation (by enabling the \texttt{ssbd} patch~\cite{ssbd}).
To derive the hardware traces, we configure \revizor{} to use two different executor modes: \texttt{TSC}~\cite{tscconfig2025microsoft}, where traces are obtained by measuring the test cases' execution time with \texttt{RDTSCP}, and \texttt{P+P}~\cite{tscconfig2025microsoft}, where hardware traces contain the cache sets that were accessed during the execution of the test case.
These modes correspond to different microarchitectural attackers: a timing-based attacker and a access-based cache attacker one~\cite{gullasch2011cache}.

\compactpara{Results}
\toolx{} performed in total 260 million test executions for each of the CPUs, over the 13 subsets of the x86 ISA and the two executor modes. The execution took 19 hours and 45 minutes per CPU on average.\footnote{During testing, \tool{}'s throughput is aligned with prior evaluations of \revizor~\cite{oleksenko2023hide}.}
\toolx{} spent an average of 
1 hours and 35 minutes per CPU
for contract refinement across all subsets, and 
6 minutes on average for contract minimization.
\Cref{tab:x86-subsets} summarizes the results of the synthesis campaign, and it highlights the types of synthesized clauses for each subset.
We manually inspected the synthesized contracts and highlight the following findings:
\begin{asparaitem}
\item \toolx{} synthesized clauses capturing the same leakage sources per subset across the different CPUs and  executor modes.\looseness=-1
\item On average, the amount of attacker indistinguishable pairs of states explored during the testing campaign with respect to the synthesized contracts is only 10.67\%.
\item For all subsets, \toolx{} synthesized clauses exposing the addresses of memory accesses, which capture leaks through the data cache. 
Even for subsets 
like \texttt{nop}, the base instructions might still result in memory operations.
E.g., \toolx{} synthesized the following clause that exposes \blueul{the address} of a \greenul{memory load/store}:
\begin{align*}
    &\mathit{cl_{mem}}
    \coloneqq\
    \blueul{\textit{OP\_VAL(0)}}
    \texttt{ IF } && \\
    &\quad [
        (
        \greenul{\textit{OP\_TYPE(0)} \texttt{ = } \texttt{mem}}
        )        
        \texttt{ and } 
        (
        \greenul{\textit{OP\_ACC(0)} \texttt{ = } \texttt{r/w}}
        )
    ]
    &&
\end{align*}
\item For the \texttt{cond} subset (the only one including control-flow statements),
\toolx{} synthesized clauses exposing the program counter, thereby exposing control-flow leaks.
For instance, \toolx{} synthesized the cluase $\mathit{cl_{pc}}$ below that exposes \blueul{the program counter} after the execution of a \greenul{control-flow statement}: %
\begin{align*}
    &\mathit{cl_{pc}}
    \coloneqq\
    \blueul{\textit{POST\_REG(\textbf{PC})}}
    \texttt{ IF } 
        (
        \greenul{\textit{OPCODE} \texttt{ = } \texttt{jmp}}
        ) 
    &\quad &\quad
    &&
\end{align*}
\item For the \texttt{dmul} subset
, \toolx{} synthesizes clauses exposing information about the divisor operand.\footnote{Following \textsc{Revizor}'s guidelines, we reduced the entropy for input generation to increase the probability of generating divisions by 0 in the tests.}
This is consistent with  prior findings showing that division operations are not constant-time~\cite{schroder2024divide}.
For instance,  \toolx{} synthesized the clause $\mathit{cl_{div}}$ exposing \blueul{the dividend} whenever \greenul{the divisor is zero}:
\begin{align*}
    &\mathit{cl_{div}}
    \coloneqq\
    \blueul{\textit{OP\_VAL(1)}}
    \texttt{ IF } && \\
    &\quad [
        (
        \textit{OPCODE} \texttt{ = } \texttt{div}
        )
        \texttt{ and }
        (
        \greenul{\textit{OP\_VAL(0)} \texttt{ = } \texttt{0}}
        ) 
    ]
    &\quad
    &&
\end{align*}
\item For the \emph{strn} subset (which includes \texttt{rep} instructions), \toolx{} synthesized clauses exposing the \texttt{rep} counter which determines how many times an instruction is repeated. As an example,
consider the following clause synthesized by \toolx{}:
\begin{align*}
    &\mathit{cl_{repe}}
    \coloneqq\
    \blueul{\textit{OP\_VAL(1)}}
    \texttt{ IF } && \\
    &\quad[
        (
        \textit{OPCODE} \texttt{ = } \texttt{repe stosd}
        ) 
        \texttt{ and }
        (
        \greenul{\textit{POST\_OP\_VAL(2)} \texttt{ = } \texttt{0}}
        ) ]
    &&
\end{align*}
Here, $\mathit{cl_{repe}}$ exposes \blueul{the destination pointer} when \greenul{the \texttt{rep} counter is zero}, i.e., in the final iteration.

\item For all target CPUs, we validated \toolx{}'s results against the instruction latencies from \textit{uops.info}~\cite{uops}, a collection of state-of-the-art reverse engineered latency models for x86 CPUs.
The leakage sources identified by \toolx{} are consistent with the \textit{uops.info}'s latency profiles, i.e., the leaks found by \toolx{} correspond to all instructions with variable latency in~\cite{uops}.

\end{asparaitem}
\smallskip

\begin{table}
    \small
    \centering
    \resizebox{\columnwidth}{!}
    {%
    \begin{tabular}{l|cccc}
    \specialrule{1.25pt}{0pt}{0pt}
    & \multirow{2}{*}{\textbf{Skylake}} & \textbf{Raptor} & \textbf{Arrow} & \textbf{Lunar}\\
    &  & \textbf{Lake} & \textbf{Lake} & \textbf{Lake}\\
    \hline \hline
    \texttt{cond:} Conditional branches        
                                            & $\blacklozenge$ $\blacksquare$ {\color{lightgray}$\bullet$} {\color{lightgray}$\blacktriangle$} %
                                            & $\blacklozenge$ $\blacksquare$ {\color{lightgray}$\bullet$} {\color{lightgray}$\blacktriangle$} %
                                            & $\blacklozenge$ $\blacksquare$ {\color{lightgray}$\bullet$} {\color{lightgray}$\blacktriangle$} %
                                            & $\blacklozenge$ $\blacksquare$ {\color{lightgray}$\bullet$} {\color{lightgray}$\blacktriangle$}\\ %
    \texttt{strn:} String operations          
                                            & $\blacklozenge$ {\color{lightgray}$\blacksquare$} $\bullet$ {\color{lightgray}$\blacktriangle$} %
                                            & $\blacklozenge$ {\color{lightgray}$\blacksquare$} $\bullet$ {\color{lightgray}$\blacktriangle$} %
                                            & $\blacklozenge$ {\color{lightgray}$\blacksquare$} $\bullet$ {\color{lightgray}$\blacktriangle$} %
                                            & $\blacklozenge$ {\color{lightgray}$\blacksquare$} $\bullet$ {\color{lightgray}$\blacktriangle$} \\ %
    \texttt{dmul:} Division and mult.         
                                            & $\blacklozenge$ {\color{lightgray}$\blacksquare$} {\color{lightgray}$\bullet$} $\blacktriangle$ %
                                            & $\blacklozenge$ {\color{lightgray}$\blacksquare$} {\color{lightgray}$\bullet$} $\blacktriangle$ %
                                            & $\blacklozenge$ {\color{lightgray}$\blacksquare$} {\color{lightgray}$\bullet$} $\blacktriangle$ %
                                            & $\blacklozenge$ {\color{lightgray}$\blacksquare$} {\color{lightgray}$\bullet$} $\blacktriangle$ \\ %
    \texttt{flag:} Operations on flags        
                                            & $\blacklozenge$ {\color{lightgray}$\blacksquare$} {\color{lightgray}$\bullet$} {\color{lightgray}$\blacktriangle$}
                                            & $\blacklozenge$ {\color{lightgray}$\blacksquare$} {\color{lightgray}$\bullet$} {\color{lightgray}$\blacktriangle$}
                                            & $\blacklozenge$ {\color{lightgray}$\blacksquare$} {\color{lightgray}$\bullet$} {\color{lightgray}$\blacktriangle$}
                                            & $\blacklozenge$ {\color{lightgray}$\blacksquare$} {\color{lightgray}$\bullet$} {\color{lightgray}$\blacktriangle$}\\
    \texttt{lock:} Atomics w/ LOCK            
                                            & $\blacklozenge$ {\color{lightgray}$\blacksquare$} {\color{lightgray}$\bullet$} {\color{lightgray}$\blacktriangle$}
                                            & $\blacklozenge$ {\color{lightgray}$\blacksquare$} {\color{lightgray}$\bullet$} {\color{lightgray}$\blacktriangle$}
                                            & $\blacklozenge$ {\color{lightgray}$\blacksquare$} {\color{lightgray}$\bullet$} {\color{lightgray}$\blacktriangle$}
                                            & $\blacklozenge$ {\color{lightgray}$\blacksquare$} {\color{lightgray}$\bullet$} {\color{lightgray}$\blacktriangle$}\\
    \texttt{atom:} Atomics w/o LOCK           
                                            & $\blacklozenge$ {\color{lightgray}$\blacksquare$} {\color{lightgray}$\bullet$} {\color{lightgray}$\blacktriangle$}
                                            & $\blacklozenge$ {\color{lightgray}$\blacksquare$} {\color{lightgray}$\bullet$} {\color{lightgray}$\blacktriangle$}
                                            & $\blacklozenge$ {\color{lightgray}$\blacksquare$} {\color{lightgray}$\bullet$} {\color{lightgray}$\blacktriangle$}
                                            & $\blacklozenge$ {\color{lightgray}$\blacksquare$} {\color{lightgray}$\bullet$} {\color{lightgray}$\blacktriangle$}\\
    \texttt{dxfr:} Data transf. (load/store)  
                                            & $\blacklozenge$ {\color{lightgray}$\blacksquare$} {\color{lightgray}$\bullet$} {\color{lightgray}$\blacktriangle$}
                                            & $\blacklozenge$ {\color{lightgray}$\blacksquare$} {\color{lightgray}$\bullet$} {\color{lightgray}$\blacktriangle$}
                                            & $\blacklozenge$ {\color{lightgray}$\blacksquare$} {\color{lightgray}$\bullet$} {\color{lightgray}$\blacktriangle$}
                                            & $\blacklozenge$ {\color{lightgray}$\blacksquare$} {\color{lightgray}$\bullet$} {\color{lightgray}$\blacktriangle$}\\
    \texttt{setc:} Conditional byte set       
                                            & $\blacklozenge$ {\color{lightgray}$\blacksquare$} {\color{lightgray}$\bullet$} {\color{lightgray}$\blacktriangle$}
                                            & $\blacklozenge$ {\color{lightgray}$\blacksquare$} {\color{lightgray}$\bullet$} {\color{lightgray}$\blacktriangle$}
                                            & $\blacklozenge$ {\color{lightgray}$\blacksquare$} {\color{lightgray}$\bullet$} {\color{lightgray}$\blacktriangle$}
                                            & $\blacklozenge$ {\color{lightgray}$\blacksquare$} {\color{lightgray}$\bullet$} {\color{lightgray}$\blacktriangle$}\\
    \texttt{nop:} NOP instructions            
                                            & $\blacklozenge$ {\color{lightgray}$\blacksquare$} {\color{lightgray}$\bullet$} {\color{lightgray}$\blacktriangle$}
                                            & $\blacklozenge$ {\color{lightgray}$\blacksquare$} {\color{lightgray}$\bullet$} {\color{lightgray}$\blacktriangle$}
                                            & $\blacklozenge$ {\color{lightgray}$\blacksquare$} {\color{lightgray}$\bullet$} {\color{lightgray}$\blacktriangle$}
                                            & $\blacklozenge$ {\color{lightgray}$\blacksquare$} {\color{lightgray}$\bullet$} {\color{lightgray}$\blacktriangle$}\\
    \texttt{logi:} Logical operations         
                                            & $\blacklozenge$ {\color{lightgray}$\blacksquare$} {\color{lightgray}$\bullet$} {\color{lightgray}$\blacktriangle$}
                                            & $\blacklozenge$ {\color{lightgray}$\blacksquare$} {\color{lightgray}$\bullet$} {\color{lightgray}$\blacktriangle$}
                                            & $\blacklozenge$ {\color{lightgray}$\blacksquare$} {\color{lightgray}$\bullet$} {\color{lightgray}$\blacktriangle$}
                                            & $\blacklozenge$ {\color{lightgray}$\blacksquare$} {\color{lightgray}$\bullet$} {\color{lightgray}$\blacktriangle$}\\
    \texttt{conv:} Data type conversion       
                                            & $\blacklozenge$ {\color{lightgray}$\blacksquare$} {\color{lightgray}$\bullet$} {\color{lightgray}$\blacktriangle$}
                                            & $\blacklozenge$ {\color{lightgray}$\blacksquare$} {\color{lightgray}$\bullet$} {\color{lightgray}$\blacktriangle$}
                                            & $\blacklozenge$ {\color{lightgray}$\blacksquare$} {\color{lightgray}$\bullet$} {\color{lightgray}$\blacktriangle$}
                                            & $\blacklozenge$ {\color{lightgray}$\blacksquare$} {\color{lightgray}$\bullet$} {\color{lightgray}$\blacktriangle$}\\
    \texttt{cmov:} Conditional moves          
                                            & $\blacklozenge$ {\color{lightgray}$\blacksquare$} {\color{lightgray}$\bullet$} {\color{lightgray}$\blacktriangle$}
                                            & $\blacklozenge$ {\color{lightgray}$\blacksquare$} {\color{lightgray}$\bullet$} {\color{lightgray}$\blacktriangle$}
                                            & $\blacklozenge$ {\color{lightgray}$\blacksquare$} {\color{lightgray}$\bullet$} {\color{lightgray}$\blacktriangle$}
                                            & $\blacklozenge$ {\color{lightgray}$\blacksquare$} {\color{lightgray}$\bullet$} {\color{lightgray}$\blacktriangle$}\\
    \texttt{bit:} Bit test and bit scan       
                                            & $\blacklozenge$ {\color{lightgray}$\blacksquare$} {\color{lightgray}$\bullet$} {\color{lightgray}$\blacktriangle$}
                                            & $\blacklozenge$ {\color{lightgray}$\blacksquare$} {\color{lightgray}$\bullet$} {\color{lightgray}$\blacktriangle$}
                                            & $\blacklozenge$ {\color{lightgray}$\blacksquare$} {\color{lightgray}$\bullet$} {\color{lightgray}$\blacktriangle$}
                                            & $\blacklozenge$ {\color{lightgray}$\blacksquare$} {\color{lightgray}$\bullet$} {\color{lightgray}$\blacktriangle$}\\
    \specialrule{1.25pt}{0pt}{0pt}
    \end{tabular}
    }
    \caption{Summary of x86-64 synthesis campaign across 13 ISA subsets and 4 Intel microarchitectures. Solid symbols indicate that \tool{} synthesized clauses capturing the corresponding leakage source ($\blacklozenge$: memory address, $\blacksquare$: control flow, $\bullet$: repetition counter, $\blacktriangle$: division by zero); grey symbols indicate that \tool{} did not synthesize such clauses.  %
    }
    \label{tab:x86-subsets}
   \vspace{0em}
\end{table}

\subsubsection{ARM}\label{sec:eval:rq3:arm}
Using \toolarm{}, we synthesize a contract associated with ARM memory instructions.
We analyzed 30 relevant load/store instructions,
including word, (signed) byte, (signed) halfword, unscaled offsets, and register pairs
(see \Cref{sec:arm-subsets} for the list of analyzed instructions).

\compactpara{Experimental setup}
We used \toolarm{} to synthesize  contracts against a Cortex-A72 (on Raspberry Pi4) and Cortex-A76 (on Raspberry Pi5) CPUs. 
To derive hardware traces, \scamv{} monitors the L1 cache state immediately after each test case execution, in an access-driven attack fashion~\cite{gullasch2011cache}.
It uses a \textit{platform module} that runs in ARM TrustZone to configure page-table attributes to control memory cacheability, clear the caches before execution, insert memory barriers around the test case code, and read the cache state using privileged debug instructions.
Each experiment was repeated 10 times, and the final cache state was checked for discrepancies.

\toolarm{} tested each instruction individually. 
For each contract refinement iteration, we used up to 10 counterexamples and 20 positive examples for synthesis.

\compactpara{Results}
The synthesis campaign took $\approx$5.5 hours for Cortex-A72 and $\approx$21.2 hours for Cortex-A76.
On average across the two campaigns, 4 minutes were spent on contract refinement and 9 minutes on minimization, with the remaining time spent generating counterexamples and positive examples.
The runtime difference arises because the Cortex-A72 tests ran in parallel on five boards, while the Cortex-A76 tests ran on a single board.

\toolarm{} produced two minimized contracts---one for each CPU---each consisting of the same six clauses.
Two clauses expose \blueul{the cache tag (all bits above bit 12) and index (bits 6–12)} associated with individual \greenul{memory loads and stores}.
\begin{align*}
    &\mathit{cl_{mem}}
    \coloneqq\
    \blueul{\textit{(OP\_VAL(1)}\texttt{[}\textit{64}\texttt{:}\textit{6}\texttt{]}\texttt{)}}
    \texttt{ IF }
        (
        \greenul{\textit{OPCODE} \texttt{ = } \texttt{ld*/st*}}
        )
    &&
\end{align*}

\toolarm successfully captures the leakage of accessed cache lines. The least significant 6 bits represent the address offset and can be ignored when identifying the accessed cache line.

The remaining four clauses are associated with \texttt{LDP}, \texttt{LDPSW}, and \texttt{STP}, which access two 32-bit words or two 64-bit doublewords from memory at the same time.
Similarly to $\mathit{cl_{mem}}$, these clauses again expose the tag and index bits of the accessed memory addresses.

\subsubsection{Summary}
These results confirm that \tool{} can synthesize contracts from black-box CPUs.
Since \tool{} relies on black-box learning, the quality of generated contracts is inherently tied to the quality of the \checker{}'s test cases, and comprehensively identifying all leaks may require more tests.
For instance, Cortex-A53 can leak the cache line offset accessed by memory loads~\cite{scamv}, but we did not observe this in our synthesis campaign.
Improving existing \checker{}s is, however, beyond the scope of our work.

\subsection{\textbf{RQ4:}  How do \tool{}-learned Contracts Compare to \textit{Template-based} Approaches?}\label{sec:eval:rq5}
\begin{table*}[t]
    \small
    \centering
    \resizebox{\textwidth}{!}{%
        \begin{tabular}{c|cccc||cccc} 
        \specialrule{1.25pt}{0pt}{0pt}
        &\begin{tabular}[c]{@{}c@{}}\textsc{\toolx}\end{tabular}
        &\begin{tabular}[c]{@{}c@{}}\textsc{LeaSyn}~\cite{wang2025synthesis}\end{tabular}
        &\begin{tabular}[c]{@{}c@{}}\textsc{RTL2$\mu$Path}~\cite{hsiao2024rtl2m}\end{tabular}
        &\begin{tabular}[c]{@{}c@{}}\textsc{VeloCT}~\cite{dinesh2025h}\end{tabular}
        &\begin{tabular}[c]{@{}c@{}}\textsc{\toolarm}\end{tabular}
        &\begin{tabular}[c]{@{}c@{}}\textsc{LeaSyn}~\cite{wang2025synthesis}\end{tabular}
        &\begin{tabular}[c]{@{}c@{}}\textsc{RTL2$\mu$Path}~\cite{hsiao2024rtl2m}\end{tabular}
        &\begin{tabular}[c]{@{}c@{}}\textsc{VeloCT}~\cite{dinesh2025h}\end{tabular}\\ \hline \hline

        \textbf{Ibex-small}             & 1.0 / 24.2 / 1.0   & 1.0 / 22.3 / 0.5  & 0.99 / 16.9 / 0.4 & 0.99 / 13.9 / 0.1 
                                        & 1.0 / 14 / 1.0     & 1.0 / 4 / 0.8     & 1.0 / 4 / 0.8     & 1.0 / 4 / 0.7   \\ \hline
        \textbf{Ibex-mult-div}          & 1.0 / 30.2 / 1.5    & 1.0 / 19.8 / 0.6   & 0.68 / 17.6 / 0.6    & 0.68 / 15.4 / 0.2 
                                        & 1.0 / 20 / 1.3     & 1.0 / 7 / 1.0      & 0.85 / 7 / 1.0       & 0.85 / 7 / 0.9  \\ \hline
        \textbf{Ibex-optimizations}     & 0.89 / 51.7 / 2.0   & 0.57 / 20.3 / 0.8  & 0.37 / 19.7 / 0.5 & 0.32 / 15.8 / 0.4
                                        & 0.92 / 35 / 4.8     & 0.56 / 17 / 3.1    & 0.52 / 13 / 2.3   & 0.52 / 14 / 2.6  \\ \hline
        \textbf{Ibex-slice}             & 0.73 / 100.5 / 2.1   & 0.41 / 29.6 / 0.4  & 0.36 / 20.3 / 0.3 & 0.18 / 15.8 / 0.2
                                        & 0.84 / 129 / 6.4     & 0.58 / 5 / 1.1     & 0.56 / 5 / 0.9    & 0.56 / 5 / 0.9  \\ \hline    
        \specialrule{1.25pt}{0pt}{0pt}
        \end{tabular}
    }
    \caption{Comparison of precision, number of clauses, and (relative) execution time with templates from previous works.}
    \label{tab:rw:precision-results}
    
\end{table*}

Previous work~\cite{DBLP:conf/sp/DineshPF24, DBLP:conf/date/MohrG024,hsiao2024rtl2m,wang2025synthesis} have developed contract synthesis tools based on templates and white-box methods which require (1) access to the processor's RTL design and (2) a set of user-provided clause templates directly defining the synthesis search space.
Our work overcomes these dependencies by introducing the first black-box synthesis approach for leakage contracts that is \textit{template-free}.
Since no existing tool operates under the same black-box, \textit{template-free} conditions as \tool{}, we cannot compare against them directly on the same hardware targets (note that \cite{DBLP:conf/sp/DineshPF24, DBLP:conf/date/MohrG024,hsiao2024rtl2m,wang2025synthesis} target small open-source RISC-V cores whereas \tool{} focuses on x86 and ARM commercial CPUs).
Instead, we design a controlled evaluation that measures the \textit{precision} of contracts synthesized by \tool{} against the same search space as each competing technique.

\compactpara{Experimental setup}
To enable a fair comparison, we define a set of ground-truth models as synthesis targets.
This allows us to do the comparison without requiring RTL access.

We define four ground-truth models inspired by the leakage profile of   the RISC-V \textbf{Ibex} core~\cite{ibex} (which is the CPU targeted by the competing \textit{template-based} tools~\cite{DBLP:conf/sp/DineshPF24, DBLP:conf/date/MohrG024,hsiao2024rtl2m,wang2025synthesis}){, and of some optimizations covering representative microarchitectural behaviors and relevant leakage scenarios (like those in \Cref{sec:eval:contracts})}:
\begin{enumerate*}[label=(\arabic*)]
    \item \textbf{Ibex-small}: the default ``small'' configuration of \textbf{Ibex} with constant-time multiplication (three cycles) and without caches {(which leaks information about the outcome of branch instructions and accessed memory addresses).}
    \item \textbf{Ibex-mult-div}: \textbf{Ibex-small} extended with a non-constant-time multiplication unit~\cite{slow-multiplier} whose execution time depends on the operand values {and it additionally leaks whether divisions and multiplications by zero happen.}
    \item \textbf{Ibex-optimizations}: \textbf{Ibex-mult-div} extended with the leaks captured by the contract models defined in \Cref{sec:eval:contracts}.
    \item \textbf{Ibex-slice}: \textbf{Ibex-mult-div} extended with memory leaks that expose only the tag and set index of memory accesses
    (\textbf{TagIdx} from \Cref{sec:eval:contracts}), and arithmetic leaks whose execution time depends on the sign (e.g., divisions) or size (e.g., multiplications) of the operands.
\end{enumerate*}
Note that \textbf{Ibex} is a RISC-V core, whose ISA is not supported by \tool{}.
For this reason, we %
instantiate the leakage profiles above for the x86 and ARM ISA (rather than RISC-V).

We compare \tool against
\textsc{LeaSyn}~\cite{wang2025synthesis} (which extends~\citet{DBLP:conf/date/MohrG024}),
\textsc{RTL2$\mu$Path}~\cite{hsiao2024rtl2m},
and \textsc{VeloCT}~\cite{dinesh2025h} (which extends~\textsc{ConjunCT}~\cite{DBLP:conf/sp/DineshPF24}),
which are the most recent and relevant \textit{template-based} white-box synthesis tools for leakage contracts.
To allow the comparison, we emulate their behavior by restricting the synthesis grammar 
to the clause templates supported by each technique.
We also run \tool{} in its original unrestricted grammar mode as an additional point of comparison.
To have a fair comparison among the tools and \tool, we run the evaluation in \toolx{} and \toolarm{} with the experimental setup as in \Cref{sec:eval:rq1:setup} (to account for the impact of different \checker{}s). 
We then measure the precision of the synthesized contracts. %
For each leakage profile and tool, we also measure the number of clauses in the final contract and the relative execution time w.r.t. \tool{}'s execution time on \textbf{Ibex-small}. %

\compactpara{Results}
\Cref{tab:rw:precision-results} summarizes the results of our analysis. 
For the simplest \textbf{Ibex-small} model, all templates are sufficient to synthesize precise contracts.
However, when considering more complex leakage profiles, restrictive templates cannot precisely express the leaks as part of a contract.
In particular, \textsc{VeloCT} (which only supports a template exposing all operands of an instruction) and \textsc{RTL2$\mu$Path} (which only supports {a template exposing a subset of all operands of an instruction})
already fail in precisely capturing the leaks of \textbf{Ibex-mult-div} since they cannot precisely capture {the specific values of the operands, e.g., $\textit{OP\_VAL(0)} \texttt{=} \texttt{0}$}.
\textsc{LeaSyn} supports more complex templates that have been hand-crafted to capture the leaks in existing \textbf{Ibex} CPUs~\cite{DBLP:conf/date/MohrG024} and, thus, its template can still precisely capture the leaks in \textbf{Ibex-mult-div}.
As soon as the templates become too restrictive, precision drops significantly.
{\tool{} can precisely capture the leaks in \textbf{Ibex-small} and \textbf{Ibex-mult-div}. Moreover, due to its template-free nature, it always results in more precise contracts than those associated with the templates supported by existing tools in complex targets like \textbf{Ibex-optimizations} and \textbf{Ibex-slice}.}
Additionally, \tool{} generates contracts with more clauses since template-freeness permits generating multiple more precise clauses. 
In contrast, other techniques operate over a more limited search space and converge earlier to a sound (but less precise) contract with fewer clauses due leakage over-approximations.
As a result, \tool{} generally incurs higher execution time, while template-based techniques are correspondingly faster due to earlier convergence.

\section{Discussion}\label{sec:discussion}
\compactpara{Limitations}
\tool{} is limited by several design choices. First, it focuses on instruction-centric contracts, synthesizing only leak clauses (what is leaked) according to the terminology of~\cite{guarnieri2021hardware,oleksenko2022revizor,ccs24}.
\tool{} also cannot represent stateful clauses that capture leaks across multiple instructions (e.g., those from computation reuse or operand packing~\cite{vicarte2021opening, ccs24}). Finally, \tool{} explores only user-mode code, as our \checker{s} do not support privileged instructions, excluding kernel-level leaks (e.g., page table accesses).

\compactpara{Black-box synthesis}
\tool{} demonstrates that leakage contract synthesis for off-the-shelf CPUs is feasible without RTL access.
RTL-based approaches~\cite{DBLP:conf/date/MohrG024,DBLP:conf/sp/DineshPF24,dinesh2025h,hsiao2024rtl2m,wang2025synthesis} can provide more exhaustive contracts but require having access to the processor's design and manual effort,
limiting their applicability. In contrast, \tool{} guarantees \emph{bounded} soundness (all detected leaks are captured), using test cases to approximate precision. This makes it practical for commercial CPUs lacking RTL or vendor-supplied specifications.

\compactpara{False negatives}
The synthesized contracts capture only those leaks that are exercised by the test cases and detected by the \checker. Leaks outside these conditions
result in \emph{false negatives}.\footnote{In \Cref{sec:evaluation}, ``false negatives'' are test cases distinguishable by hardware but not by contract traces; thus 100\% soundness means capturing all \emph{detected} leaks, not all possible leaks.} 

\compactpara{False positives}
Despite 
minimization (\Cref{sec:approach}), synthesized contracts may still over-approximate actual leaks, resulting in \emph{false positives}. This occurs if the \postprocessor{} cannot remove all over-approximations, or if some leaks (e.g., when only a hash of a register leaks) are inherently hard to infer precisely from test cases alone.

\compactpara{Threats to validity of the evaluation}
To cover a large space of test cases and to reduce biases introduced by manually selected test-cases, we decided to construct the validation sets using the random program generators included in the two relational testing tools supported by \tool{}, that is, \revizor{} and \scamv{}.
In particular, for \toolx, on average, more than 70\% of the trace pairs are distinguishable by the target, whereas for \toolarm, the validation set (generated using the SMT-backend of \scamv) is constructed so that 50\% of the trace pairs are distinguishable.
This ensures that the generated validation sets exercise both leaking and non-leaking behaviors, and, thus, result in informative metrics.

\compactpara{Other side channels}
\tool{} generalizes to other observational models (e.g., power or encrypted memory) by integrating domain-specific checkers. For instance, power leaks can be modeled via ISA-level observational equivalence~\cite{DBLP:conf/ccs/BloemGGHMP22}, requiring a checker that can test for power leaks and find positive/negative examples. 

\compactpara{Leakage contracts and secure programming}
Contracts provide the foundations for leak-free software.
As shown in \cite{guarnieri2021hardware}, ensuring that leakage traces are not dependent on program secrets is sufficient to guarantee the absence of microarchitectural secret-dependent leaks for any CPU satisfying the contract.
Since our leakage clauses are of the form $\mathit{ex}\ \mathtt{IF}\ \mathit{pr}$, one needs to ensure that (a) the predicate $\mathit{pr}$ is always secret-independent and that (b) whenever $\mathit{pr}$ holds, then the expression $\mathit{ex}$ is also secret-independent.

\section{Related Work}\label{sec:relatedwork}
Over the years, many approaches have been developed to analyze and mitigate hardware side-channels. 
Here, we summarize works related to leakage contracts and existing techniques to extract them. We also briefly review works on different synthesis approaches, contract synthesis, and automating side-channel analysis.

\compactpara{Leakage Contracts}
Several works aim to formalize microarchitectural features to bridge the hardware-software abstraction gap~\cite{DBLP:conf/sp/CauligiDMBS22}. Some introduce operational semantics for speculative and out-of-order execution~\cite{guarnieri2020spectector,xaverccs,patrignani2021exorcising,cauligi2020constant,DBLP:conf/ccs/GuancialeBD20,vassena2021automatically,mcilroy2019spectre,daniel2021hunting}, at varying abstraction levels. Others capture microarchitectural side-effects using axiomatic semantics~\cite{mosier2022axiomatic,disselkoen2018code,DBLP:conf/sp/LeonK22}. Our work builds directly on the leakage contract framework~\cite{guarnieri2021hardware}, which provides a formal basis for relating contracts to concrete leaks in CPUs.

\compactpara{Synthesis of Contracts}
The closest work to \tool{} includes RTL2M$\mu$PATH~\cite{hsiao2024rtl2m}, which verifies contracts directly from RTL, and Mohr et al.~\cite{DBLP:conf/date/MohrG024} and \textsc{LeaSyn}~\cite{wang2025synthesis}, who build contracts from manually-specified clauses and RTL tests. \textsc{ConjunCT}~\cite{DBLP:conf/sp/DineshPF24}, VeloCT~\cite{dinesh2025h} synthesize coarse-grained contracts for timing attacks, classifying instructions as ``safe'' or ``unsafe'' from RTL. We remark that all these tools are template-based, that is, they require  user-provided clause templates directly constraining the synthesis search space. In contrast, \tool{} automatically learns precise, fine-grained contracts for black-box CPUs without templates or RTL access, using counter\-example-guided synthesis.

\compactpara{Counterexample-Guided Synthesis}
Different works have applied counterexample-Guided Synthesis (CEGIS)~\cite{alur2013syntax} to various tasks, including those related with the hardware-software interface.
For instance,  Heule et al.~\cite{heule2016stratified} and Liu et al.~\cite{liu2024synthesizing} apply CEGIS to automatically synthesize semantics and language semantics.
In contrast, PipeSynth~\cite{norman2023pipesynth} uses formal synthesis to generate axioms for microarchitectural memory consistency. 
Finally, \textsc{SynthCT}~\cite{dinesh2022synthct} synthesizes translations of safe/unsafe instructions.
\tool{}' contract synthesis follows at a high-level the standard CEGIS approach applying example-guided synthesis to leakage contracts expanding the space of examples to pairs of traces instead of single examples.

\compactpara{Automatic Reasoning about Side Channels}
Various fra\-meworks automate side-channel analysis, including the fuz\-zers used in this work, \revizor{}~\cite{oleksenko2022revizor,oleksenko2023hide} and \scamv{}~\cite{scamv,scamv:micro}. Other tools include Osiris~\cite{DBLP:journals/corr/abs-2106-03470} (timing-based fuzzing using ISA specs), CheckMate~\cite{checkmate} (automated exploit/test generation for hardware), and Xenon~\cite{v2021solver} (formal verification of constant-time hardware). There are  other approaches that leverage symbolic execution, fuzzing, or formal analysis for side-channel vulnerability detection~\cite{DBLP:conf/uss/MoghimiLS020,DBLP:journals/corr/abs-2106-03470,oleksenko:specfuzz,vassena2021automatically,wang2020kleespectre,yang2023pensieve,wang2023specification}.\looseness=-1

\section{Conclusion}\label{sec:conclusions}
A precise ISA-level leakage contract is critical for secure programming. We introduced a method to extract such contracts from black-box CPUs and validated it with \tool{} on \texttt{x86} and \texttt{ARM} CPUs.
Our results show that \tool{} can synthesize precise and boundedly sound contracts that capture all exercised instruction-level microarchitectural leaks.
Furthermore, the results show that our \textit{template-free} approach eliminates the inflexibility of conventional methods, leading to more precise contracts than \textit{template-based} approaches.
\vspace{1em}

\begin{acks}
This work has been partially supported
by the Wallenberg AI, Autonomous Systems and Software Program (WASP) funded by the Knut and Alice Wallenberg Foundation,
by the \grantsponsor{1}{European Union}{https://erc.europa.eu/} under the ERC project \grantnum{1}{Primula 101230068},\footnote{Views and opinions expressed are however those of the author(s) only and do not necessarily reflect those of the European Union or the European Research Council. Neither the European Union nor the European Research Council can be held responsible for them.} 
by the \grantsponsor{3}{Spanish Ministry of Science and Innovation}{https://www.ciencia.gob.es/} under the Ram\'on y Cajal grant \grantnum{3}{RYC2021-032614-I}, 
by the \grantsponsor{3}{Spanish Ministry of Science and Innovation}{https://www.ciencia.gob.es/} under the project \grantnum{3}{PID2022-142290OB-I00 ESPADA}, 
by the \grantsponsor{3}{Spanish Ministry of Science and Innovation}{https://www.ciencia.gob.es/} under the project \grantnum{3}{CEX2024-001471-M}, 
by the \grantsponsor{3}{Spanish Ministry of Science and Innovation}{https://www.ciencia.gob.es/} under the project \grantnum{3}{EUR2025-164828 SIN TRAZAS}, 
and by a  gift from Intel and Amazon.\looseness=-1
\end{acks}

\bibliographystyle{ACM-Reference-Format}
\bibliography{refs}

\appendix
\section{x86 Instruction Subsets Used in \Cref{sec:eval:rq3:x86} }\label{sec:x86-subsets}

The synthesis campaign in \Cref{sec:eval:rq3:x86} targets 13 subsets of the x86-64 ISA.
These subsets have been derived from a prior Revizor testing campaign~\cite{oleksenko2023hide}.
Altogether, these subsets result in a complete base x86-64 (user-level) instruction set, excluding those instructions that are not (yet) compatible with \revizor.
The subsets do not cover: 
\begin{inparaenum}[(a)]
\item system instructions,
\item instruction incorrectly emulated by Unicorn, 
\item control-flow instructions not supported by Revizor (e.g., indirect jumps), and
\item ISA extensions such as AVX or x87 that require more complex test case generation algorithms to avoid failures, which are also not compatible with \revizor.
\end{inparaenum}

Each of the tested subsets consisted of several basic arithmetic instructions (including their versions with memory operands) and of several instructions that are unique to the given subset. The exact instructions are as follows:

\begin{itemize}
  \item \texttt{cond} (conditional branches): \texttt{ADC}, \texttt{ADD}, \texttt{CMP}, \texttt{DEC}, \texttt{INC}, \texttt{NEG}, \texttt{SBB}, \texttt{SUB}, \texttt{J*}, \texttt{LOOP*}, \texttt{JMP} (unconditional direct jump).
  
  \item \texttt{strn} (string operations): \texttt{ADC}, \texttt{ADD}, \texttt{CMP}, \texttt{DEC}, \texttt{INC}, \texttt{NEG}, \texttt{SBB}, \texttt{SUB}, \texttt{CLC}, \texttt{CLD}, \texttt{CMC}, \texttt{LAHF}, \texttt{LOCK}, \texttt{REPE}, \texttt{REPNE}, \texttt{SAHF}, \texttt{SCASB}, \texttt{SCASD}, \texttt{SCASW}, \texttt{STC}, \texttt{STD}.
  
  \item \texttt{dmul} (division and multiplication): \texttt{ADC}, \texttt{ADD}, \texttt{CMP}, \texttt{DEC}, \texttt{INC}, \texttt{NEG}, \texttt{SBB}, \texttt{SUB}, \texttt{DIV}, \texttt{IMUL}, \texttt{MUL}.

  \item \texttt{flag} (operations on flags): \texttt{ADC}, \texttt{ADD}, \texttt{CMP}, \texttt{DEC}, \texttt{INC}, \texttt{NEG}, \texttt{SBB}, \texttt{SUB}, \texttt{CLC}, \texttt{CLD}, \texttt{CMC}, \texttt{LAHF}, \texttt{SAHF}, \texttt{STC}, \texttt{STD}.                                                                                                                                                                                                                                

  \item \texttt{lock} (atomics with \texttt{LOCK} prefix): \texttt{ADC}, \texttt{ADD}, \texttt{CMP}, \texttt{DEC}, \texttt{INC}, \texttt{NEG}, \texttt{SBB}, \texttt{SUB}, \texttt{LOCK ADC}, \texttt{LOCK ADD}, \texttt{LOCK CMP}, \texttt{LOCK DEC}, \texttt{LOCK INC}, \texttt{LOCK NEG}, \texttt{LOCK SBB}, \texttt{LOCK SUB}, \texttt{LOCK BSF}, \texttt{LOCK BSR}, \texttt{LOCK BT}, \texttt{LOCK BTC}, \texttt{LOCK BTR}, \texttt{LOCK BTS}, \texttt{LOCK NOT}, \texttt{LOCK OR}, \texttt{LOCK TEST}, \texttt{LOCK XOR}.
  
  \item \texttt{atom} (atomics without \texttt{LOCK} prefix): \texttt{ADC}, \texttt{ADD}, \texttt{CMP}, \texttt{DEC}, \texttt{INC}, \texttt{NEG}, \texttt{SBB}, \texttt{SUB}, \texttt{CMPXCHG}, \texttt{XADD}, \texttt{LOCK CMPXCHG}, \texttt{LOCK XADD}.
  
  \item \texttt{dxfr} (data transfer): \texttt{ADC}, \texttt{ADD}, \texttt{CMP}, \texttt{DEC}, \texttt{INC}, \texttt{NEG}, \texttt{SBB}, \texttt{SUB}, \texttt{BSWAP}, \texttt{MOV}, \texttt{MOVSX}, \texttt{MOVZX}, \texttt{XCHG}.
  
  \item \texttt{setc} (conditional byte set): \texttt{ADC}, \texttt{ADD}, \texttt{CMP}, \texttt{DEC}, \texttt{INC}, \texttt{NEG}, \texttt{SBB}, \texttt{SUB}, \texttt{SET*}.
  
  \item \texttt{nop} (\texttt{NOP} instructions): \texttt{ADC}, \texttt{ADD}, \texttt{CMP}, \texttt{DEC}, \texttt{INC}, \texttt{NEG}, \texttt{SBB}, \texttt{SUB}, \texttt{NOP}.
  
  \item \texttt{logi} (logical operations): \texttt{ADC}, \texttt{ADD}, \texttt{CMP}, \texttt{DEC}, \texttt{INC}, \texttt{NEG}, \texttt{SBB}, \texttt{SUB}, \texttt{AND}, \texttt{NOT}, \texttt{OR}, \texttt{TEST}, \texttt{XOR}.
  
  \item \texttt{conv} (data type conversion): \texttt{ADC}, \texttt{ADD}, \texttt{CMP}, \texttt{DEC}, \texttt{INC}, \texttt{NEG}, \texttt{SBB}, \texttt{SUB}, \texttt{CBW}, \texttt{CDQ}, \texttt{CWD}, \texttt{CWDE}.
  
  \item \texttt{cmov} (conditional moves): \texttt{ADC}, \texttt{ADD}, \texttt{CMP}, \texttt{DEC}, \texttt{INC}, \texttt{NEG}, \texttt{SBB}, \texttt{SUB}, \texttt{CMOV*}.
  
  \item \texttt{bit} (bit test and bit scan): \texttt{ADC}, \texttt{ADD}, \texttt{CMP}, \texttt{DEC}, \texttt{INC}, \texttt{NEG}, \texttt{SBB}, \texttt{SUB}, \texttt{BSF}, \texttt{BSR}, \texttt{BT}, \texttt{BTC}, \texttt{BTR}, \texttt{BTS}.
\end{itemize}

\section{ARM Instruction Set Used in \Cref{sec:eval:rq3:arm}}\label{sec:arm-subsets}

The synthesis campaign from \Cref{sec:eval:rq3:arm} targets a subset of the memory instructions in the ARM ISA.
These instructions cover various addressing modes and data sizes, including standard, unprivileged, and unaligned variants. The selected subset ensures representative coverage of typical memory access patterns relevant to our analysis.
Next, we list all memory instructions covered by our campaign: 

\begin{itemize}
\item \texttt{LDR}, \texttt{LDRB}, \texttt{LDRH}, \texttt{LDRSB}, \texttt{LDRSH}, \texttt{LDRSW}

Load instructions for various data sizes (byte, halfword, signed byte/halfword, word).

\item \texttt{STR}, \texttt{STRB}, \texttt{STRH}

Store instructions for byte, halfword, and word sizes.

\item \texttt{LDUR}, \texttt{LDURB}, \texttt{LDURH}, \texttt{LDURSB}, \texttt{LDURSH}, \texttt{LDURSW}

Load instructions with unscaled immediate offset addressing.

\item \texttt{STUR}, \texttt{STURB}, \texttt{STURH}

Store variants corresponding to \texttt{LDUR} instructions.

\item \texttt{LDP}, \texttt{LDPSW}, \texttt{STP}

Load/store pair instructions for loading/storing two registers simultaneously.

\item \texttt{LDTR}, \texttt{LDTRB}, \texttt{LDTRH}, \texttt{LDTRSB}, \texttt{LDTRSH}, \texttt{LDTRSW}

Load instructions with unprivileged access supporting various data sizes and signed variants.

\item \texttt{STTR}, \texttt{STTRB}, \texttt{STTRH}

Store instruction variants with unprivileged access.
\end{itemize}

\section{\tool Evaluation of Parameters}\label{sec:x86_best_config}
In this section, we evaluated the impact of different parameters of \tool using \toolx.
In particular, we evaluated the impact of: number of positive examples (\Cref{sec:x86_best_config:pex}), contract resetting parameter $R$ (\Cref{sec:x86_best_config:resetting}),synthesis depth (\Cref{sec:x86_best_config:synthDepth}), and minimization options (\Cref{sec:x86_best_config:minimization}).

\subsection{Impact of Positive Examples}\label{sec:x86_best_config:pex}
To address leakage over-approximations, \tool{} uses positive examples to guide the \refiner{} about which executions should not be distinguished by the synthesized clause. 

\compactpara{Experimental setup}
For each target ground-truth contract, we randomly generate 100 programs and 100 random states per program, and we use \toolx to synthesize the corresponding 
contracts for different numbers of positive examples, ranging from 0 (no positive examples) to 150.
Afterwards, we measure the precision of the synthesized contracts and the average execution time of clause refinement step (i.e., the average execution time of a single call to the \refiner).

\begin{figure}[h] 
  \includegraphics[width=\columnwidth]{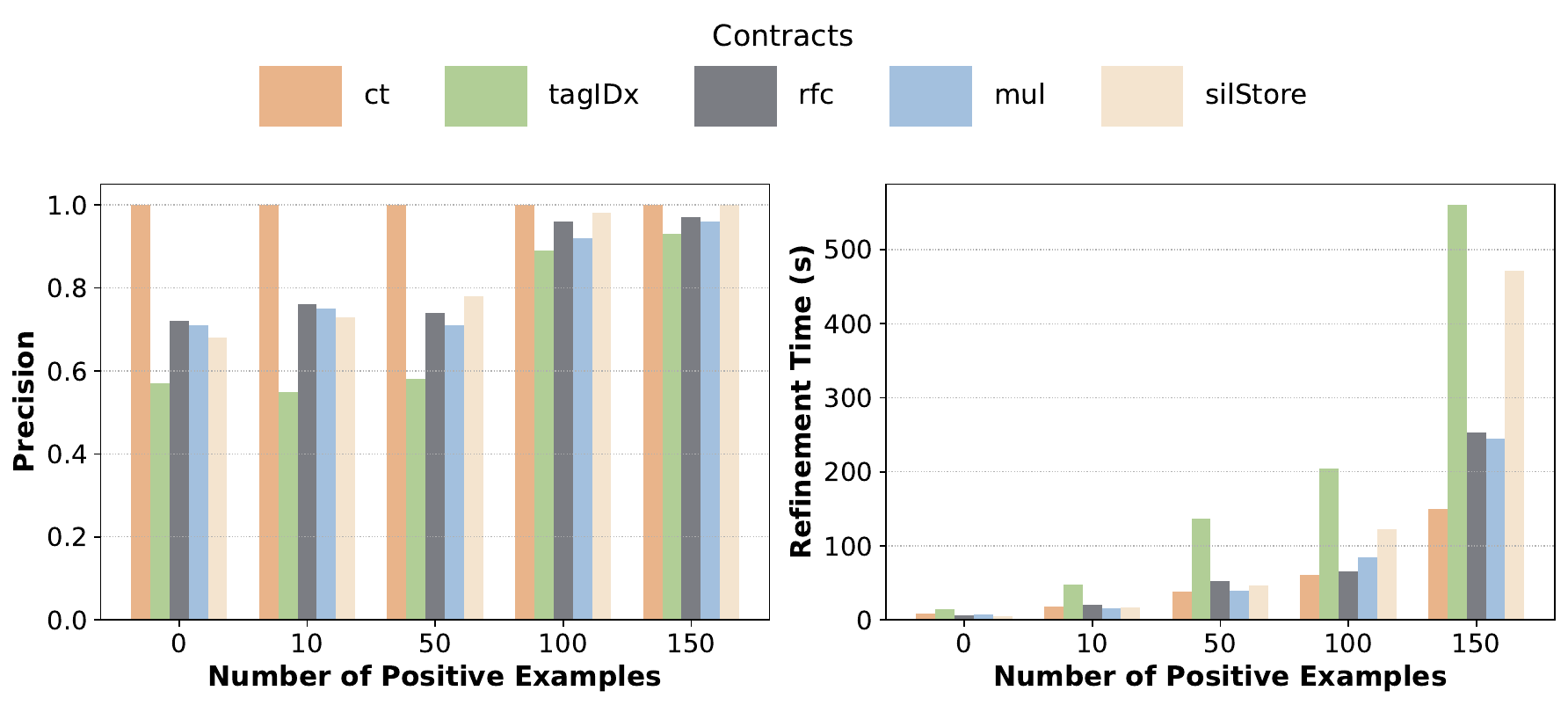} 
  \caption{\toolx{}: Impact of positive examples on precision and refinement time.}
  \label{fig:x86-pex}
\end{figure}

\compactpara{Results}
\Cref{fig:x86-pex} shows the impact of the number of positive examples on the  precision of the synthesized contract (plot on the left) and on the  clause refinement time (plot on the right).
Note that both plots report the average metrics computed across all 100 programs.
We highlight that increasing the number of positive examples increases both the precision of the synthesized contracts as well as the refinement time (since more positive examples result in more complex synthesis constraints).
We remark, however, that the precision starts to plateau at 100 positive examples, whereas the refinement time keeps increasing.
For this reason, we decided to use 100 positive examples in the evaluation in \Cref{sec:evaluation} as it is the best trade-off between achieved precision and refinement time.

\subsection{Impact of Contract Resetting}\label{sec:x86_best_config:resetting}
\begin{table*}[t]
    \small
    \centering
    \begin{tabular}{c|c|c|c|c|c|c|c}
        \specialrule{1.25pt}{0pt}{0pt}
        \multicolumn{8}{c}{\textbf{\textsc{\toolx}}} \\
        \hline \hline
        Contract & \begin{tabular}[c]{@{}c@{}}N. Programs\\ Before Restart\end{tabular} &
    \begin{tabular}[c]{@{}c@{}}N. Clauses\\ Before Unification\end{tabular} &
    \begin{tabular}[c]{@{}c@{}}N. Clauses\\ After Unification\end{tabular} &
    \begin{tabular}[c]{@{}c@{}}N. Clauses\\ After Minim.\end{tabular} &
    \begin{tabular}[c]{@{}c@{}}avg. Contract\\ Minim. Time\end{tabular} &
    \begin{tabular}[c]{@{}c@{}}avg. Total\\ Refinement Time\end{tabular} &
    \begin{tabular}[c]{@{}c@{}}avg. \\ P/S\end{tabular} \\
        \hline

        \multirow{5}{*}{ct} 
        & 100     & 2826.8 & 1093.9 & 6 & 15h 34min 45s & 26h 36min 47s  & 1.0/1.0\\
        & 500     & 365.8  & 288.2  & 6 &  54min 17s    & 5h 40min 47s   & 1.0/1.0\\
        & 1000    & 170.7  & 154.8  & 6 &  14min 40s    & 3h 4min 44s    & 1.0/1.0\\
        & 5000    & 87.3   & 35.8   & 5 &   1min 20s    &  27min 25s     & 1.0/1.0\\
        & 10000   & 19.6   & 19.4   & 4.2 &     30s       & 13min 58s      & 0.99/1.0\\
        \hline

        \multirow{5}{*}{TagIdx} 
        & 100     & 3005.4 & 1285.7  & 8.4  & 6h 38min 51s & 54h 34min 35s  & 0.96/1.0\\
        & 500     & 407.3  & 388.9   & 14.4 & 1h 18min 46s & 12h 49min 47s  & 0.79/1.0\\
        & 1000    & 243.9  & 240   & 17.1 & 14min 18s    & 7h 14min 26s   & 0.35/1.0\\
        & 5000    & 156.2  & 57    & 11.3 &  1min 9s     & 1h 47min 25s   & 0.10/0.99\\
        & 10000   & 30.3   & 30.3    & 7.3  &   31s        & 56min 22s      & 0.10/0.99\\
        \hline

        \multirow{5}{*}{RFC} 
        & 100     & 3117.3 & 1126.7 & 9.8 & 14h 11min 14s & 24h 16min 53s & 1.0/1.0\\
        & 500     & 496.3  & 368.8  & 9.8 & 1h 5min 2s    & 6h 7min 23s   & 1.0/1.0\\
        & 1000    & 245.2  & 209.1  & 11 & 21min 4s      & 3h 28min 17   & 0.88/1.0\\
        & 5000    & 102  & 56.1   & 11.5 & 2min 28s      & 36min 58s     & 0.59/1.0\\
        & 10000   & 31.1   & 31.1   & 7.2  & 51s           & 18min 59s     & 0.48/1.0\\
        \hline

        \multirow{5}{*}{mul} 
        & 100     & 3788.1 & 1422.4 & 16 & 19h 55min 59s & 57h 38min 43s  & 1.0/1.0\\
        & 500     & 485.4  & 255.5  & 15.7 & 3h 17min 4s   & 13h 29min 35s  & 1.0/1.0\\
        & 1000    & 319.3  & 285  & 16.6 & 44min 3s      & 7h 28min 29s   & 0.99/1.0\\
        & 5000    & 122.9  & 73.8   & 13.2 & 3min 44s      & 1h 45min 19s   & 0.41/1.0\\
        & 10000   & 40.9   & 40.9   & 10.4 & 1min 17s      & 1h 3min 4s     & 0.41/1.0\\
        \hline

        \multirow{5}{*}{SilStore} 
        & 100     & 304.4 & 105.6 & 1 & 14min 27s & 3h 48min 40s & 1.0/1.0\\
        & 500     & 47.2  & 32.2  & 1 & 2min 3s   & 1h 13min 48s & 1.0/1.0\\
        & 1000    & 22.3  & 18.3  & 1.5 & 38s       & 33min 37s    & 0.83/1.0\\
        & 5000    & 12.2  & 3.9   & 1.7 & 7s        & 4min 21s     & 0.49/1.0\\
        & 10000   & 2.7   & 2.2   & 2.2 & 3s        & 1min 45s     & 0.43/1.0\\
        \specialrule{1.25pt}{0pt}{0pt}
    \end{tabular}
    \caption{The table shows the impact on the final contract size (in clauses), minimization time, total refinement time, and average precision and soundness metrics given different values of the resetting parameter $R$ (i.e., after how many tested programs we reset the contract candidate to $\emptyset$).}
    \label{tab:contract-reset}
\end{table*}

Next, we study the impact of the contract resetting parameter $R$ (which indicates after how many programs is the candidate contract reset as indicated in \Cref{sec:synthesis:initial}) on the contract synthesized by \toolx{}.

\compactpara{Experimental setup}
We use \toolx{} to learn 
contracts for all the ground truth contracts as targets. 
For each ground-truth contract, we use 10000 programs with 100 states each, and 100 positive examples. 
We ran \toolx{} with different values for the resetting parameter $R$ ranging from 100 (so, resetting after every 100 programs) to 10000 (so, no resets during synthesis).
We repeat the experiment outlined above 10 times, with different initial randomness seeds.
Each time, we measure the number of clauses before and after post-processing, the total execution time for the contract minimization and clause refinement (i.e., the sum of all time spent in the \refiner), and the average precision and soundness of the synthesized contracts against a validation set of 50000 programs with 100 inputs each.\looseness=-1

\compactpara{Results}
\Cref{tab:contract-reset} reports the results of this experiment.
We highlight the following insights: 
\begin{compactenum}[(1)]
    \item A lower value of $R$ (i.e., more frequent restarts  from an empty candidate contract) result in more precise and sound contracts.
    Intuitively, this happens because restarts allow \toolx{} to ``recover'' from over-approximated clauses synthesized in previous steps.
    \item More frequent restarts result in a higher execution time. 
    This is due to the fact that synthesis from an empty contract often requires exploring more counterexamples (and, therefore, performing more calls to the synthesizer), thereby increasing \toolx{}'s execution time.    
\end{compactenum}
In the evaluation in \Cref{sec:evaluation}, we decided to set $R$ to 500 as it offers a good trade-off between total execution time and contract precision.

\subsection{Impact of Synthesis Depth}\label{sec:x86_best_config:synthDepth}
\begin{figure}[t] 
  \includegraphics[width=\columnwidth]{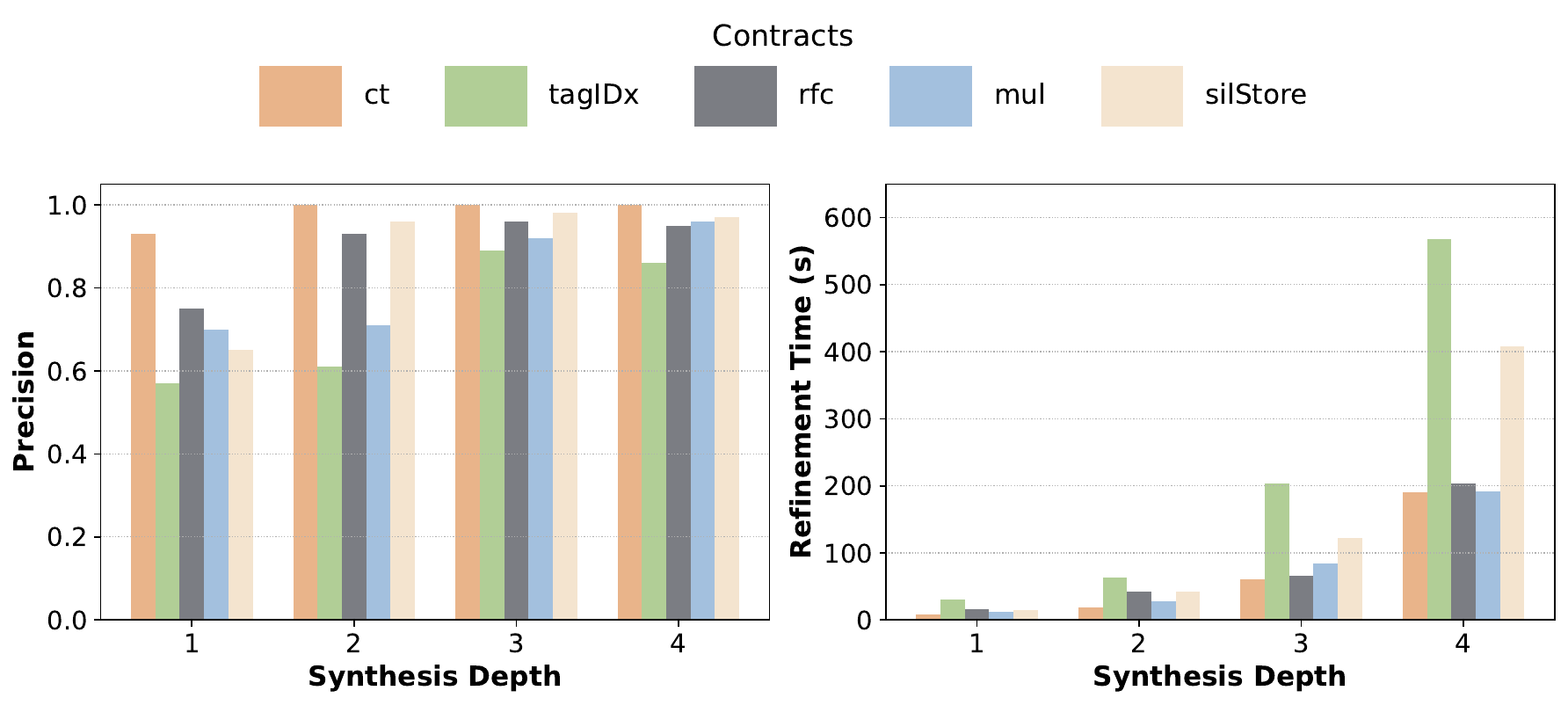} 
  \caption{\toolx{}: Impact of synthesis depth on precision and refinement time.}
  \label{fig:x86-depth}
  \vspace{-1.2em}
\end{figure}

The synthesis depth limits the complexity of the \icl clauses explored by the synthesizer, where synthesis depth \textit{n} indicates that the solver will consider only expressions whose syntax tree has depth \textit{n}. A small depth value restricts the ability of the solver to find complex solutions by limiting the search space to ``simple'' expressions. Conversely, an excessive depth exponentially increases the search space and the synthesis time.

\compactpara{Experimental setup}
For each target ground-truth contract, we randomly generate 100 programs and 100 random states per program, and we use \toolx to synthesize the corresponding 
contracts for different synthesis depth values, ranging from 1 to 4.
Afterwards, we measure the precision of the synthesized contracts and the average execution time of clause refinement step (i.e., the average execution time of a single call to the \refiner).

\compactpara{Results}
\Cref{fig:x86-depth} shows the impact of synthesis depth  on the (average) precision of the synthesized contract (plot on the left) and on the (average)  clause refinement time (plot on the right).
Note that both plots report the average metrics computed across all 100 programs.
We highlight that increasing the synthesis depth improves both the precision of synthesized contracts and the refinement time.
However, precision starts to plateau at depth $3$, whereas refinement time keeps increasing (due to a larger search space).
In particular,  for synthesis depth 4, the solver reaches a time-out (set at 600 seconds) in 89 of the 100 cases of the \textbf{TagIdx} contract.
For this reason, we decided to use synthesis depth 3 in the evaluation in \Cref{sec:evaluation} as it is the best trade-off between precision and refinement time.

\begin{table*}[t]
    \small
    \centering
    \begin{tabular}{c|cc|cc|cc|cc}
        \specialrule{1.25pt}{0pt}{0pt}
        \multicolumn{9}{c}{\textbf{\textsc{\toolx}}} \\
        \hline \hline
        Contract  & \multicolumn{2}{c|}{All Minimization} & \multicolumn{2}{c|}{Ctr. Minimization} & \multicolumn{2}{c|}{TC Minimization} & \multicolumn{2}{c}{No Minimization} \\
        \cline{2-9}
        & \multicolumn{1}{c|}{avg. P/S} & avg. Time & \multicolumn{1}{c|}{avg. P/S} & avg. Time & \multicolumn{1}{c|}{avg. P/S} & avg. Time & \multicolumn{1}{c|}{avg. P/S} & avg. Time \\
        \hline

        ct
        & \multicolumn{1}{c|}{1.0/1.0} & 11h 57min & \multicolumn{1}{c|}{1.0/1.0} & 29h 17min & \multicolumn{1}{c|}{0.70/1.0} & 10h 31min & \multicolumn{1}{c|}{0.68/1.0} & 25h 30min  \\
        \hline

        TagIdx
         & \multicolumn{1}{c|}{0.79/1.0} & 18h 57min & \multicolumn{1}{c|}{0.30/1.0} & 33h 48min & \multicolumn{1}{c|}{0.14/1.0} & 14h 14min & \multicolumn{1}{c|}{0.06/1.0} & 33h 7min  \\
        \hline

        RFC
         & \multicolumn{1}{c|}{1.0/1.0} & 10h 32min & \multicolumn{1}{c|}{0.92/1.0} & 28h 11min & \multicolumn{1}{c|}{0.30/1.0} & 8h 40min & \multicolumn{1}{c|}{0.27/1.0} &  24h 41min  \\
        \hline

        mul
        & \multicolumn{1}{c|}{1.0/1.0} & 24h 21min & \multicolumn{1}{c|}{0.91/1.0} & 51h 5min & \multicolumn{1}{c|}{0.20/1.0} &14h 58min & \multicolumn{1}{c|}{0.19/1.0} & 47h 40min  \\
        \hline

        SilStore
         & \multicolumn{1}{c|}{1.0/1.0} & 1h 23min & \multicolumn{1}{c|}{1.0/1.0} & 10h 15min & \multicolumn{1}{c|}{0.13/1.0} & 1h 19min & \multicolumn{1}{c|}{0.07/1.0} & 10h 7min  \\        
        \specialrule{1.25pt}{0pt}{0pt}
    \end{tabular}
    \caption{Precision (P), Soundness (S) and average of total time for different minimization configurations on 
    contracts.}
    \label{tab:minimizers-x86}
\end{table*}
\subsection{Impact of Minimization}\label{sec:x86_best_config:minimization}

Next, we study the impact of the three kinds of minimization optimizations used by \toolx---program/data minimization (from Revizor, \Cref{sec:implementation:oracle}) and contract minimization (\Cref{sec:implementation:post-processing}).

\compactpara{Experimental setup}
We use \toolx to learn a 
contract for all the ground truth contracts as target and using 10000 programs with 100 states each and resetting after 500 programs.
We ran \toolx{} using four minimization configurations: 

\begin{compactenum}[(1)]
    \item \emph{All Minimization}: all minimization optimizations are enabled;
    \item \emph{Ctr. Minimization}: only contract minimization is enabled;
    \item \emph{TC Minimization}: only program/data minimization is enabled;
    \item \emph{No Minimization}: no minimization optimizations are enabled.
\end{compactenum}

We repeat all the experiments outlined above 10 times, with different initial randomness seeds.
For each configuration, we measured the total execution time of \toolx{}  as well as the average precision and soundness of the synthesized contracts against a validation set of 50000 programs with 100 inputs each.

\compactpara{Results}
\Cref{tab:minimizers-x86} reports our experiment results. 
We highlight the following findings:
\begin{compactenum}
    \item Enabling all minimization optimizations increases precision by an average of 70\% with respect to the \emph{No Minimization} cases. %
    \item The execution time is mostly affected by the use of program/data minimization. Not using program/data minimization
    results in more complex constraints (and a larger search space) for the solver, which again results in longer execution times (see, for instance, how configurations \emph{All Minimization} and \emph{Prg. Minimization} are faster than configurations \emph{Ctr. Minimization} and \emph{No Minimization}). 
    \item Contract minimization is the optimization that contributes the most to improving precision, as it eliminates overapproximations in the final contracts.
\end{compactenum}
Based on these results, in the evaluation in \Cref{sec:evaluation} we always enable all three optimizations.

\end{document}